\documentclass[screen,acmsmall]{acmart}

\AtBeginDocument{%
  \providecommand\BibTeX{{%
    \normalfont B\kern-0.5em{\scshape i\kern-0.25em b}\kern-0.8em\TeX}}}

\setcopyright{acmcopyright}

\acmJournal{TOSEM}
\acmVolume{}
\acmNumber{}
\acmArticle{}
\acmMonth{0}

\graphicspath{{./figs/}}
\DeclareGraphicsExtensions{.pdf,.jpeg,.png,.eps}

\usepackage{caption}
\usepackage{subcaption}
\usepackage{booktabs, tabularx}
\usepackage{threeparttable}
\usepackage{longtable}

\usepackage{algorithm}
\usepackage{algpseudocode}
\floatname{algorithm}{Procedure}
\usepackage{enumerate}
\usepackage{fancybox}  
\usepackage{multirow}
\usepackage{pifont}
\usepackage{soul}
\usepackage[table,xcdraw]{xcolor}
\usepackage{minted}

\usepackage{listings}
\usepackage{makecell}
\usepackage{float}
\usepackage[export]{adjustbox}
\usepackage{enumitem}
\definecolor{codegreen}{rgb}{0,0.6,0}
\definecolor{codegray}{rgb}{0.5,0.5,0.5}
\definecolor{codepurple}{rgb}{0.58,0,0.82}

\lstdefinestyle{mypython}{
    language=Python,
    commentstyle=\color{codegreen},
    keywordstyle=\color{magenta},
    numberstyle=\tiny\color{codegray},
    stringstyle=\color{codepurple},
    basicstyle=\ttfamily\footnotesize,
    breakatwhitespace=false,
    breaklines=true,
    captionpos=b,
    keepspaces=true,
    numbers=left,
    numbersep=5pt,
    showspaces=false,
    showstringspaces=false,
    showtabs=false,
    tabsize=4
}

\newcommand{\secref}[1]{Section~\ref{#1}}
\newcommand{\tabref}[1]{Table~\ref{#1}}

\DeclareTextFontCommand{\emp}{\bfseries}

\usepackage[most]{tcolorbox}
\definecolor{custom-gray}{cmyk}{0, 0, 0, 0.7, 1.00}
\newtcolorbox{Summary}[2][]{
top=0.15in,
fonttitle=\bfseries,
colbacktitle=custom-gray,
colback=gray!5,
colframe=gray!40!black,
enhanced,
attach boxed title to top left={xshift=1.5em,yshift=-\tcboxedtitleheight/2},
boxed title style={size=small,colback=custom-gray},
drop shadow={black!50!white},
title=#2,#1}

\definecolor{mygreen}{RGB}{0,158,115}
\definecolor{myyellow}{RGB}{230,159,0}
\definecolor{myorange}{RGB}{255, 87, 37}
\definecolor{mycyan}{RGB}{0, 173, 181}
\definecolor{myblue}{RGB}{0, 114, 178}
\sethlcolor{yellow}
\newcommand{\gopifix}[1]{}
\newcommand{\haofix}[1]{}

\begin{document}

\title[Studying the Security and Maintainability of MCP Servers]{Model Context Protocol (MCP) at First Glance:\\ Studying the Security and Maintainability of MCP Servers}

\author{Mohammed Mehedi Hasan}
\orcid{0000-0001-9837-0998}
\email{mohammedmehedi.hasan@queensu.ca}
\affiliation{%
  \institution{Queen's University}
  \city{Kingston}
  \state{ON}
  \country{Canada}
  \postcode{K7L 1A8}
}
\author{Hao Li}
\orcid{0000-0003-4468-5972}
\affiliation{%
  \institution{Queen's University}
  \city{Kingston}
  \state{ON}
  \country{Canada}
  \postcode{K7L 1A8}
}
\email{hao.li@queensu.ca}

\author{Emad Fallahzadeh}
\orcid{0009-0005-5024-4868}
\affiliation{%
  \institution{Queen's University}
  \city{Kingston}
  \state{ON}
  \country{Canada}
  \postcode{K7L 1A8}
}
\email{emad.fallahzadeh@queensu.ca}
\author{Gopi Krishnan Rajbahadur}
\orcid{0000-0003-1812-5365}
\affiliation{%
  \institution{Queen's University}
  \department{School of Computing}
  \city{Kingston}
  \state{Ontario}
  \country{Canada}
}
\email{16gkr1@queensu.ca}

\author{Bram Adams}
\orcid{0000-0001-7213-4006}
\affiliation{%
  \institution{Queen's University}
  \city{Kingston}
  \state{ON}
  \country{Canada}
  \postcode{K7L 1A8}
}
\email{bram.adams@queensu.ca}

\author{Ahmed E. Hassan}
\orcid{0000-0001-7749-5513}
\affiliation{%
  \institution{Queen's University}
  \city{Kingston}
  \state{ON}
  \country{Canada}
  \postcode{K7L 1A8}
}
\email{ahmed@cs.queensu.ca}

\renewcommand{\shortauthors}{M. Mehedi Hasan et al.}

\begin{abstract}

Although Foundation Models (FMs), such as GPT-4, are increasingly used in domains like finance and software engineering, reliance on textual interfaces limits these models' real-world interaction. To address this, FM providers introduced tool calling---triggering a proliferation of frameworks with distinct tool interfaces. In late 2024, Anthropic introduced the Model Context Protocol (MCP) to standardize this tool ecosystem.  With SDK downloads surpassing twenty five million per week and 86\% of enterprises using models supporting MCP tools, MCP is rapidly emerging as a de facto industry standard.  Despite its adoption, MCP's AI-driven, non-deterministic control flow introduces new risks to sustainability, security, and maintainability, warranting closer examination.

Towards this end, we present the first large-scale empirical study of MCP. Using state-of-the-art health metrics and a hybrid analysis pipeline, combining a general-purpose static analysis tool with an MCP-specific scanner, we evaluate 1,899 open-source MCP servers to assess their health, security, and maintainability. Despite MCP servers demonstrating strong health metrics, we identify eight distinct vulnerabilities---only three overlapping with traditional software vulnerabilities. Additionally, 7.2\% of servers contain general vulnerabilities and 5.5\% exhibit MCP-specific tool poisoning. Regarding maintainability, while 66\% exhibit code smells, 14.4\% contain ten bug patterns overlapping prior research. These findings highlight the need for MCP-specific vulnerability detection techniques while reaffirming the value of traditional analysis and refactoring practices. Furthermore, we advocate for stronger governance across the MCP ecosystem by incorporating MCP-specific vulnerabilities into standardized vulnerability databases, enabling automated security scanning within MCP registries, and promoting responsible development practices to ensure the long-term safety and sustainability of the MCP ecosystem.

\end{abstract}




\maketitle
\newcommand{\rqzero}{How healthy and sustainable are MCP servers?}
\newcommand{\rqone}{To what extent do MCP servers contain security vulnerabilities?}
\newcommand{\rqtwo}{To what extent do MCP servers contain maintainability issues?}

\newcommand{\motivation}{\emp{Motivation. }}
\newcommand{\approach}{\medskip\noindent\emp{Approach. }}
\newcommand{\findings}{\medskip\noindent\emp{Findings. }}
\newcommand{\runningexample}{\textbf{Running example. }}


\section{Introduction}\label{sec:introduction}
Foundation Models (FMs), such as GPT-4~\cite{openai2024gpt4} and LLaMA~\cite{touvron2023llama}, have revolutionized Artificial Intelligence (AI) applications through their powerful capabilities of language understanding, generation, and reasoning. Their widespread adoption is evident as they are increasingly integrated into real-world products and applications across domains, e.g., finance~\cite{wu2023bloomberggpt}, healthcare~\cite{denecke2024potential}, and software development~\cite{weber2024significant}, fundamentally transforming how we interact with technology and information. Despite this momentum, FMs inherently operate through isolated ``textual'' interfaces, limiting their ability to directly interact with dynamic, real-world data sources, APIs, and systems. 

As an initial step to bridge this gap, FM providers have introduced capabilities such as function calling~\cite{openai-function-calling, llama-function-calling} and tool calling~\cite{claude-tool-calling}, commonly known as \textit{tool calling}, enabling real-world interactions---for example, booking an appointment using calendar tools. Leveraging these capabilities, AI and Agent frameworks, e.g., LangChain~\cite{langchain}, Autogen~\cite{autogen}, Pydantic-AI~\cite{pydantic-ai}, CrewAI~\cite{crew-ai}, Dify~\cite{dify-ai} and LlamaIndex~\cite{llamaindex}, have introduced FM-based tool workflows for building AI applications that can reason about tasks, plan tool usage, and coordinate tool invocation. However, a significant challenge has emerged from this proliferation: each framework typically defines its own tool interface and data structure, e.g., a tool defined in one AI framework is not directly portable in another AI framework, leading to fragmentation that hampers interoperability, scalability, and consistent security practices.

To address these limitations, Anthropic introduced the Model Context Protocol (MCP)~\cite{anthropic_mcp}---a universal, client-server protocol standardizing how AI applications expose tools to FMs. Analogous to how package managers (e.g., NPM, PyPI~\cite{bommarito2019empirical}) standardized software component reuse across diverse domains, MCP provides a unified, open standard allowing compliant clients to discover and invoke tools from a diverse ecosystem of MCP servers, thereby enhancing interoperability in AI tool usage.

In the broader context of AI and software engineering, MCP serves as a foundational element of AI-native application architecture by enhancing interoperability and reducing ecosystem fragmentation~\cite{hou2025model}. Since its introduction in late November 2024, major stakeholders such as Microsoft~\cite{microsoft_mcp}, OpenAI~\cite{openai_mcp}, Google~\cite{google_mcp}, and Cloudflare~\cite{cloudflare_mcp} have all embraced MCP to support AI applications. For instance, Microsoft's Azure MCP server\footnote{\url{https://github.com/Azure/azure-mcp}} allows FMs to interact with Azure services, e.g., cloud storage and databases. By providing Foundation Models (FMs) with a standardized mechanism to interact with external systems, MCP eliminates the need to rewrite  custom tool interfaces for specific models or for agent-specific frameworks. This relieves the practitioners from both FM and framework vendor lock-in, while also enabling faster development cycles through the reuse of existing MCP servers. Hence the widespread adoption of MCP has fueled its growth, with the official MCP Software Development Kit surpassing twenty five million weekly downloads ~\cite{mcp-npm, mcp-pip} and giving an estimated 86 percent of enterprises access to MCP-capable workflows through models that support tool calling\footnote{\url{https://menlovc.com/wp-content/uploads/2025/11/menlo_ventures_mid-year_llm_report-2025.pdf}}. This swift uptake by industry leaders and strong community engagement through open-source channels highlight MCP's growing centrality and further motivate the need to examine MCP servers through the lens of open-source software concerns such as health, sustainability, vulnerability, and maintainability.

Similar to other reusable software packages deployed in production environments, e.g., PyPI and NPM packages, the health, sustainability, security, and maintainability of MCP servers are critical to ensure long-term growth. Experience from other software ecosystems suggests that communities prioritize secure, widely adopted, well-documented packages~\cite{mujahid2023characteristics}. However, early studies have demonstrated that FMs can be coerced into utilizing tools from vulnerable MCP servers to compromise user systems, commonly known as \texttt{tool poisoning}, leading to attacks such as credential theft and malicious code execution ~\cite{narajala2025enterprise, hou2025model}. For example, an MCP server can have a tool with file write permission, and by manipulating the FM, it can be tricked into adding some malicious code snippets in the \texttt{.bashrc} file to open a backdoor port for attackers~\cite{radosevich2025mcp}. Given that the whole point of MCP server tools is to give FMs autonomy in deciding when and which tool to use, the presence of potential security (and maintainability) issues in said tools has raised concerns among industry stakeholders~\cite{microsoft-mcp-security}. 

These concerns are not merely theoretical; they have already manifested in real-world security incidents. For instance, the recent discovery of the first malicious Postmark-MCP server, which silently added a hidden BCC (Blind Carbon Copy) and exfiltrated thousands of user emails, illustrates how a compromised MCP server can bypass conventional security controls and leak sensitive data at scale~\footnote{\url{https://thehackernews.com/2025/09/first-malicious-mcp-server-found.html}}. In parallel, the security community is identifying critical vulnerabilities in the developer supply chain of MCP servers. A recent example is the critical remote-code-execution vulnerability (CVE-2025-49596) in MCP Inspector, a popular debug tool, which allows arbitrary command execution on developers’ machines simply by visiting a malicious website\footnote{\url{https://nvd.nist.gov/vuln/detail/CVE-2025-49596}}. These incidents underscore the urgent need to examine MCP server health, security, and maintainability with empirical rigor, since flaws in MCP servers can directly compromise user systems, data integrity, and supply-chain trust.

However, despite these concerns, the current state of publicly available open-source MCP servers remains unexplored, particularly in key software dimensions such as health, sustainability, security, and maintainability. Understanding these dimensions is critical because early-stage development health and sustainability can directly influence the trustworthiness of MCP servers and determine whether they can be reliably adopted in practice. Moreover, without visibility into their security posture or maintainability characteristics, practitioners risk deploying systems without understanding their actual attack surface or long-term operational implications. These gaps are concerning, given the potential impact of unaddressed software issues. For instance, historically, we know that one Heartbleed vulnerability alone caused an estimated USD \$500 million in damages globally~\cite{heartbleed-cost} in 2014, and recent industrial reports suggest that mature organizations may spend up to £250,000 (USD~\$332,500) per system annually on software maintenance~\cite{state_of_software_2025}. Without rigorous empirical analysis of these three dimensions, the community lacks the measures to mitigate such risks and issues proactively. 

Based on this motivation, we distill three open questions that currently lack empirical evidence in the MCP ecosystem: (i) how actively is the MCP ecosystem maintained; (ii) how prevalent are traditional and MCP-specific vulnerabilities in publicly available MCP servers and how does this landscape different from other software ecosystems; and (iii) whether maintainability issues, e.g., code smells and bugs, raise concerns comparable to those observed in mature ecosystems and thereby hinder adoption. To bridge this gap, we conduct the first large-scale empirical study on health, sustainability, security vulnerabilities and maintainability issues across 1,899 MCP servers. To date (Jun 1, 2025) our dataset consists of 343 MCP servers from the official MCP collection~\cite{mcp_official_collection} complemented by 1,556 MCP servers mined from open-source repositories hosted on GitHub. By analyzing these MCP servers, we aim to answer the following Research Questions (RQs):

\begin{enumerate}[label=\textbf{RQ-\arabic*:}, start=0]

\item \emp{\rqzero}\\
\motivation As MCP is becoming a crucial part of the AI application ecosystem, it is essential to understand how healthy and sustainable the MCP ecosystem currently is. Prior studies have reported that different development (commits, CI adoption, etc.) and community metrics (contributor count, contributor reputation, issue activity, etc.) can act as indicators of health and long-term sustainability~\cite{goggins2021open,xiao2023early, coelho2017modern}. As development metrics reflect continuous maintenance activity and community metrics capture stakeholder engagement, we can understand the current state-of-the-art health and sustainability characteristics of MCP servers by studying these metrics and interpreting them in contrast to general OSS  and Machine Learning(ML) software metrics.  

\emp{Findings.} MCP servers demonstrate strong development and community metrics with an average of \textbf{5.5 commits/week} (vs. traditional software 2.5 commits/week) and a slightly \textbf{higher CI adoption rate} (42.2\% vs. traditional 40.3\%), comparable median contributor count (2.0 vs. 2.0), but \textbf{significantly higher social reach} in terms of median follower count of the contributors (\textbf{129.6 followers} vs. traditional 37.3), and a slightly slower median issue resolution time (5.6 days vs. traditional 4.0 days).  

\item \emp{\rqone}

\motivation Prior studies have reported that at least 46\% of Python packages contain security vulnerabilities~\cite{ruohonen2021large} and the \textbf{average cost of some security incidents, e.g., data breach, can be USD up to \$4.5 million}~\cite{ibm-breach-cost}. The potential risk may be even higher in the context of MCP servers, as these systems bridge foundation models with databases, file systems, and APIs, with humans handing off control to FM models. To understand the actual risks, we need to characterize the vulnerability patterns of MCP servers, analyze those against the known traditional software engineering vulnerabilities, and evaluate the effectiveness of current vulnerability detection tools in identifying MCP-specific issues. 

\emp{Findings.} To explore the vulnerability landscape of MCP servers, we apply a traditional vulnerability detector, SonarQube, and identify that \textbf{7.2\% of MCP servers are affected by eight distinct vulnerability patterns}, with credential exposure being the most prevalent (3.6\%). Only three of those vulnerabilities overlap with known ecosystem vulnerabilities. However, by applying an MCP-specific scanning tool~\cite{mcp-scan}, we uncover \textbf{5.5\% MCP servers suffer from tool poisoning}, which indicates that traditional tools only find part of the whole landscape and we need more robust MCP-specific vulnerability detection techniques and tools.

\item \emp{\rqtwo}

\motivation Prior studies have reported that maintainability issues, e.g., code smells, can be linked to bugs or fault-proneness ~\cite{palomba2018diffuseness}. Software development teams can spend up to 39\% of maintenance effort on fixing these in mature projects~\cite{amit2021corrective}. As an emerging new technology, the maintainability landscape of MCP servers, e.g., code smells and bugs, is still unknown to the community. To understand the impact of these maintainability issues on MCP servers, we need to characterize the code smells and bugs and analyze those against the known traditional software engineering issues to evaluate the long-term impact. 

\emp{Findings.} MCP servers exhibit \textbf{similar code smell and bug patterns} like traditional software engineering domain. In the long term, we suspect these code smells, and bugs can incur significant maintainability issues in the MCP domain, as observed in traditional software engineering~\cite{state_of_software_2025}. We report a taxonomy of ten code smells and nine bugs in MCP code. We also find that \textbf{66\% MCP servers} have critical or blocker-level code smells, where \textbf{high cognitive complexity is the most prevalent smell} and \textbf{14.4\% MCP servers have critical or blocker level-bugs}.

\end{enumerate}

The main contributions of our paper are as follows:
\begin{enumerate}

\item \textbf{Assets:}
    \begin{enumerate}
        \item \textbf{Dataset:} We release a curated dataset of \textbf{1{,}899 open-source MCP servers}, including 88 official, 255 community-maintained, and 1{,}556 GitHub-mined servers obtained through a structured discovery pipeline. This unified dataset represents \textbf{the first full-spectrum coverage of the MCP ecosystem} and enables reproducible, extensible, and longitudinal ecosystem-level research, serving as a foundational dataset for future MCP-specific studies.

        \item \textbf{Comparative Baselines:} We establish cross-ecosystem baselines for evaluating the health, sustainability, security, and maintainability of MCP servers. We compile and normalize reference metrics from established open-source software engineerin and machine learning ecosystems, including PyPI, NPM, and Infrastructure-as-Code (IaC), to create baselines that integrate health, sustainability, vulnerability, and code smell indicators. These baselines enable consistent and transferable benchmarking of emerging ecosystems against mature open-source domains.
    \end{enumerate}

\vspace{2ex}

\item \textbf{Method:}

We formulate a hybrid analysis framework that integrates general-purpose static analysis with MCP-specific vulnerability detection. To the best of our knowledge, this is \textbf{the first framework that complements a traditional static analyzer such as SonarQube with an MCP-specific scanner (\texttt{mcp-scan})}, enabling joint evaluation of both traditional and MCP-specific vulnerabilities within a single empirical pipeline. This framework provides a \textbf{reusable and extensible foundation} for analyzing other emerging software engineering domains by substituting domain-specific scanners and targets, thereby enabling scalable, reproducible, and comparative studies of vulnerability and maintainability across ecosystems.

\vspace{2ex}

\item \textbf{Empirical Findings:}

We evaluate the health, security, and maintainability of MCP servers through large-scale empirical analysis. Our results show that MCP servers demonstrate strong health and sustainability metrics but still contain both traditional and MCP-specific vulnerabilities, along with frequent maintainability issues such as code smells and recurring bug patterns.

\end{enumerate}

The remainder of this paper is outlined as follows. Section~\ref{sec:motivation}  provides a motivating example for MCP. Section~\ref{sec:background} provides background information. Section~\ref{sec:relatedwork} gives an overview of related work. Section~\ref{sec:exp_design} describes the design of our study. Section~\ref{sec:results} presents the results. Section~\ref{sec:implications} discusses the implications of our findings. Section~\ref{sec:threadstovalidity} outlines threats to the validity of our study, and Section~\ref{sec:conclusion} concludes the paper.

\section{A Motivational Example}\label{sec:motivation}
Developing real-world AI applications with FMs demands seamless orchestration of diverse components. To motivate the role of MCP and it's challenges studied in this work, we present a common scenario faced by AI application developers while integrating FMs in the real world. 

Alex is an AI engineer responsible for building and maintaining FM-based AI applications. They have been tasked with developing a conversational shopping assistant that lets users chat about products, receive recommendations, and complete purchases via Stripe.
Alex begins by exploring whether a single FM can handle the entire flow---from product discovery to payment---but quickly realizes that:

\begin{figure}[t]
    \centering

    \begin{subfigure}[b]{0.32\textwidth}
        \centering
        \includegraphics[width=\textwidth, frame]{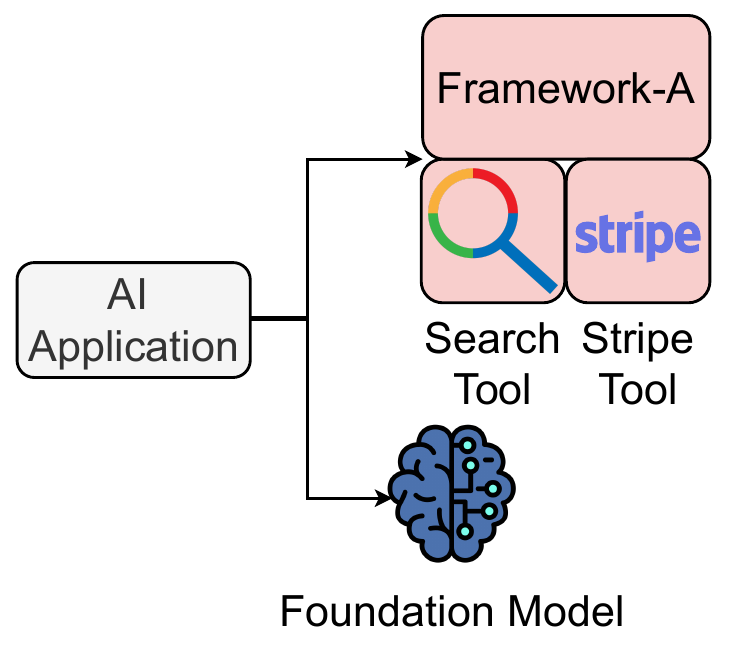}
        \caption{AI Application developed with framework A using built-in search and stripe tool.}
        \label{fig:motivating-example-langchain}
    \end{subfigure}
    \hfill
    \begin{subfigure}[b]{0.32\textwidth}
        \centering
        \includegraphics[ width=\textwidth, frame]{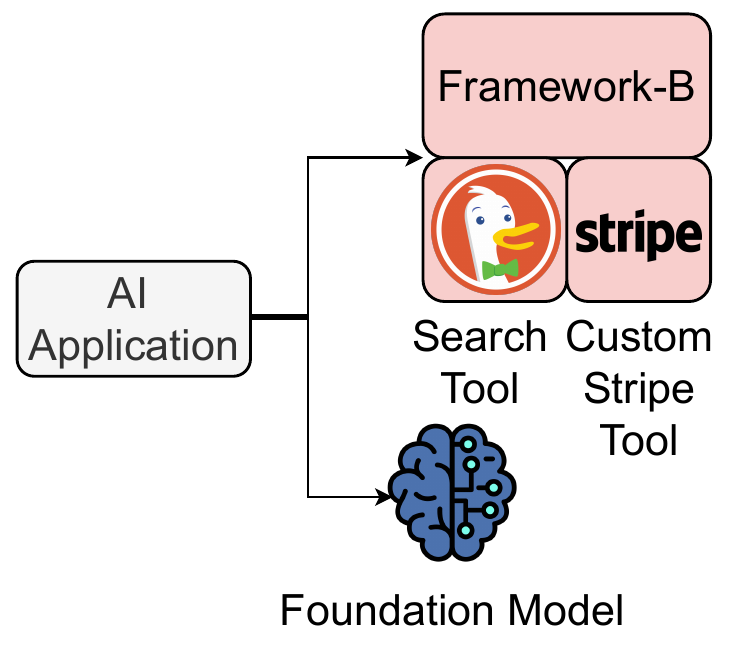}
        \caption{A's tool is incompatible with framework B, needing to implement a custom stripe tool.}
        \label{fig:motivating-example-llamaindex}
    \end{subfigure}
    \hfill
    \begin{subfigure}[b]{0.32\textwidth}
        \centering
        \includegraphics[width=\textwidth, frame]{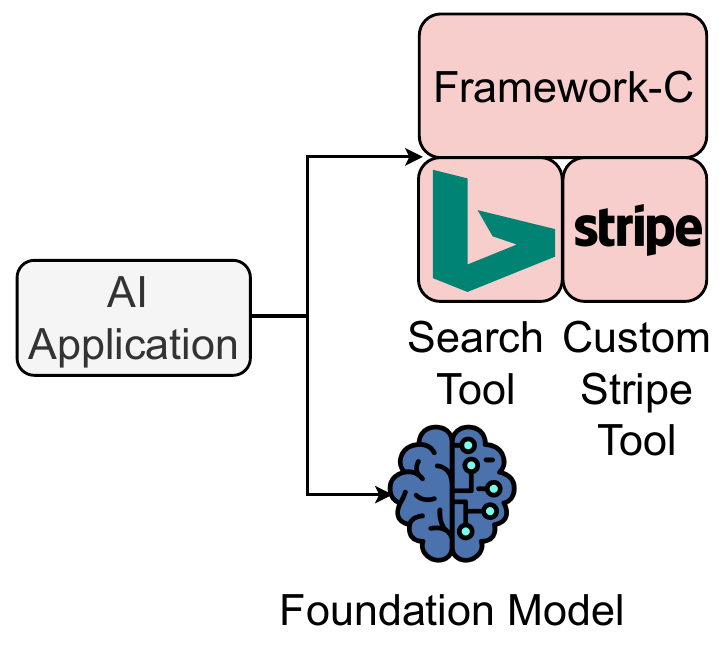}
        \caption{Custom stipe tool made for framework B is incompatible with framework C.}
        \label{fig:motivating-example-pydantic}
    \end{subfigure}

    \vspace{1em}

    \includegraphics[ width=\textwidth]{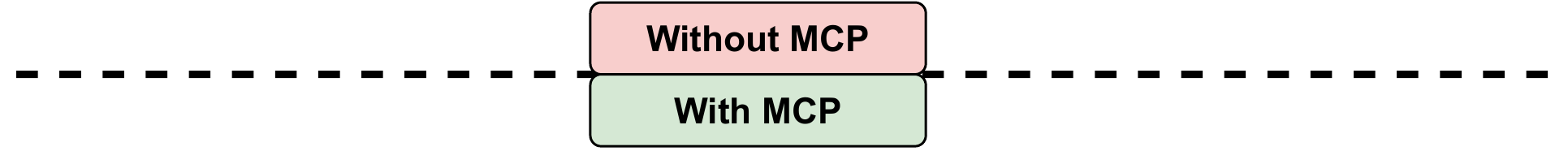}

    \vspace{1em}

    \begin{subfigure}[b]{0.99\textwidth}
        \centering
        \includegraphics[width=\textwidth, frame]{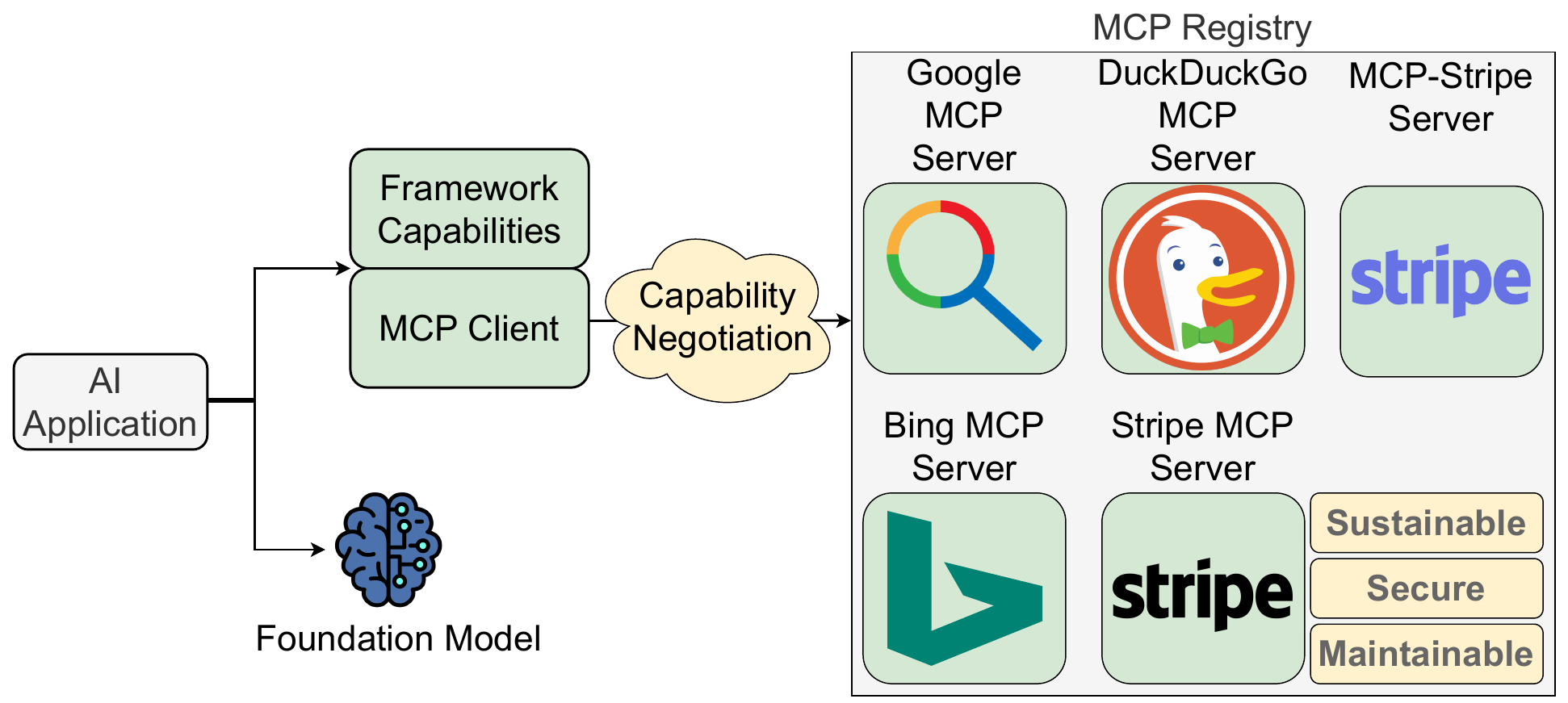}
        \caption{MCP: decoupling tools, prompts and resources from frameworks}
        \label{fig:motivating-example-mcp}
    \end{subfigure}

    \caption{A motivating example of developing FM-based AI applications. In (a), Alex developed an AI application using framework A and did not need any custom tools. In (b), when enhancing an existing application written in framework B, they had to build a custom stripe tool because B does not support A’s built-in tools. In (c), they had to re-implement the same stripe tool again with a different interface to integrate it into the C framework. In contrast (d), where MCP servers offer a way to decouple tools from frameworks and enable interoperability — but raise new questions around sustainability, security, and maintainability.}
    \label{fig:motivating-example-v2}
\end{figure}

\begin{enumerate}
    \item \textbf{Finding products} requires a dedicated search mechanism (e.g., web search APIs).
    \item \textbf{Recommending products} needs an LLM to synthesize and rank the discovered products
    \item \textbf{Collecting payments} needs a payment gateway, e.g., Stripe integration.
\end{enumerate}

Alex learned they need \texttt{tools}, a self-contained implementation of some functionalities that connect a model to existing APIs or data sources, to accomplish 1 and 3. After exploring several AI application frameworks that can coordinate different tools, Alex opted for framework A (for example, LangChain) as it has an extensive library of pre-built tools, including various search integrations (e.g., Google Search, Bing Search) and, importantly, a tool for Stripe integration. 

\textbf{Scenario 1: The Good.} Using framework A, Alex builds an AI application to orchestrate a user purchase workflow. When a user expresses a requirement, the system: (1) sends the query along with available tools (e.g., search and Stripe) to the FM; (2) invokes the search tool when the FMs instructs to do so (e.g., with keywords, brands, sites) and returns the results; (3) presents the FM's synthesized results to the user and collects the product choice; and (4) executes the Stripe tool, based on the FM's instruction to complete the payment.

\textbf{Scenario 2: The Bad.} The shopping assistant quickly gained popularity, particularly its integrated payment feature, which encouraged a new requirement: integrating Stripe-based donation collection to an existing fundraising AI application built with coordination framework B, for example LlamaIndex. As B did not have a Stripe tool, Alex attempted to reuse the A's Stripe tool but found it incompatible; A's tool interfaces, invocation methods, and data handling mechanisms fundamentally differed from B
's expectation. Consequently, Alex ended up re-implementing a custom Stripe tool for B.

\textbf{Scenario 3: The Ugly.}
Soon after, another team required to add Stripe payments to their AI analytics app built with framework C, e.g., Pydantic-AI, to collect premium membership fees. Again, no built-in tool existed. Alex attempted to adapt their custom B Stripe tool, but C's tool interface prevented reuse, forcing them to consider a third bespoke Stripe implementation.

Through this series of experiences, Alex realized the critical root problem: a fundamental lack of tool interoperability and standardization within the FM-based AI application ecosystem. 

\textbf{Scenario 4: The Promise.} To address this, Alex explores the emerging Model Context Protocol (MCP), which standardizes how FMs discover and invoke external capabilities via standard interfaces without any hard coupling with the framework behind the AI application. After MCP’s release, Alex finds several Stripe MCP servers from open-source community. They realize that they can now integrate these directly into any of their applications, as long as the applications leverage the MCP protocol, finally eliminating duplicated effort and code.

\textbf{Scenario 5: The Challenge.} However, payment processing is highly sensitive for both customers and Alex's company. As a revenue-generating application, they require the Stripe MCP server to be sustainable, e.g., regularly developed and bug-fixed. Moreover, they must guard against security vulnerabilities compromising sensitive customer data (e.g., credit card numbers) and the maintenance issues of downstream applications. Given the availability of multiple Stripe MCP servers from various communities, they are now perplexed about how to determine:
\begin{enumerate}
    \item Which MCP servers are the most healthy and sustainable for their critical AI applications? Alex knows that the long-term viability of their applications depends on whether the underlying MCP servers evolve reliably. Without understanding this, they risk choosing an MCP server that may fail to keep pace with real-world operational demands. But he cannot find any study to understand the current health and sustainability of MCP servers, which we address in our \textbf{RQ-0}.
    \item How can they verify an MCP server's security to protect sensitive customer data (e.g., credentials, credit card details)? Alex is aware that MCP servers may contain vulnerabilities, but they are unclear about how often such issues appear in the public servers they rely on or how to detect them effectively. This uncertainty makes Alex concerned that they might choose an MCP server without understanding its true security landscape. We address this concern in our \textbf{RQ-1}.
    \item How can they assess an MCP server's maintenance quality, such as its likelihood of having bugs or smells? Alex worries that these maintainability issues could introduce significant technical debt and refactoring costs, but they cannot find any reference or study that describes what the maintainability profiles of MCP servers look like, which we address in our \textbf{RQ-2}.
\end{enumerate}
In this study we want to explore these challenges through quantative and qualitative analysis.
\section{Background}\label{sec:background}
\subsection{Tool Landscape for FM-based AI Applications}
\subsubsection{Tools for FM}
Foundation models (FMs) are pre-trained on massive datasets to internalize broad linguistic patterns and world knowledge. They rely entirely on their statically encoded parameters (weights) at inference time and cannot access information beyond their pre-training data~\cite{brown2020language, bommasani2021opportunities}. Though Retrieval-Augmented Generation (RAG) augments FMs with domain-specific textual context~\cite{lewis2020retrieval}, FMs are still confined to generating only text output. To interact with the real world, FM-based AI applications use tools, which are external interfaces performing operations on behalf of the model, ranging from web search, database queries, and API invocation to code execution and hardware control. These primitives enable AI applications to fetch real-time data, interact with user interfaces, and trigger real-world actions. However, setting up those tools for proper use can be challenging and time-consuming~\cite{hassan2024rethinking}.

\subsubsection{AI Frameworks}
Several tool-centric frameworks have emerged in recent years to support the development of FM-based AI applications and AI agents. Examples include LangChain~\cite{langchain}, Autogen~\cite{autogen}, and LlamaIndex~\cite{llamaindex}. These frameworks provide infrastructure, abstractions, and orchestration mechanisms for building AI applications and agents to reason about tasks, plan tool usage, and coordinate tool invocation to pursue user-defined goals.

\subsubsection{Tool Workflow}

Most frameworks offer built-in tools and provide support for creating custom tools to enable AI applications to specify a step-by-step tool workflow. As illustrated in our motivating example  Figure~\ref{fig:motivating-example-langchain}, once the user expresses an “intent,” the AI application formats the request, retrieves the available tools, and bundles them into a prompt for the FM. The FM then plans an execution path and guides the AI application through a series of tool calls and interactions to fulfill the tasks, e.g., using a search tool to find products, collecting user preferences via interaction, and calling a payment API through Stripe tool to complete the purchase journey.  

\subsubsection{Shortcomings of Framework-Specific Tools}
While these frameworks support FM-tool integration, they also introduce significant interoperability challenges due to their siloed operation~\cite{hassan2024rethinking}. Each framework defines its schema for tool registration, invocation, and response formatting, leading to compatibility issues. For example, a Stripe tool implemented for LangChain may not function seamlessly with AutoGen or LlamaIndex because of differences in input specifications and data contracts. Moreover, tools are often tightly coupled with their originating frameworks and lack standardized discoverability mechanisms, making reuse difficult. So, the application built on a specific framework is limited to using the tools provided only by that framework and misses the opportunity to integrate any other outside tools. This fragmentation mirrors early challenges in software engineering, which were eventually addressed through standardized packaging systems like NPM~\cite{zerouali2018empirical} or PyPI~\cite{priyalakshmi2018evaluation}, enabling consistent reuse, interoperability, and lifecycle management.

\subsection{Model Context Protocol(MCP)}

\begin{figure}[th]
    \centering
    \includegraphics[width=\textwidth]{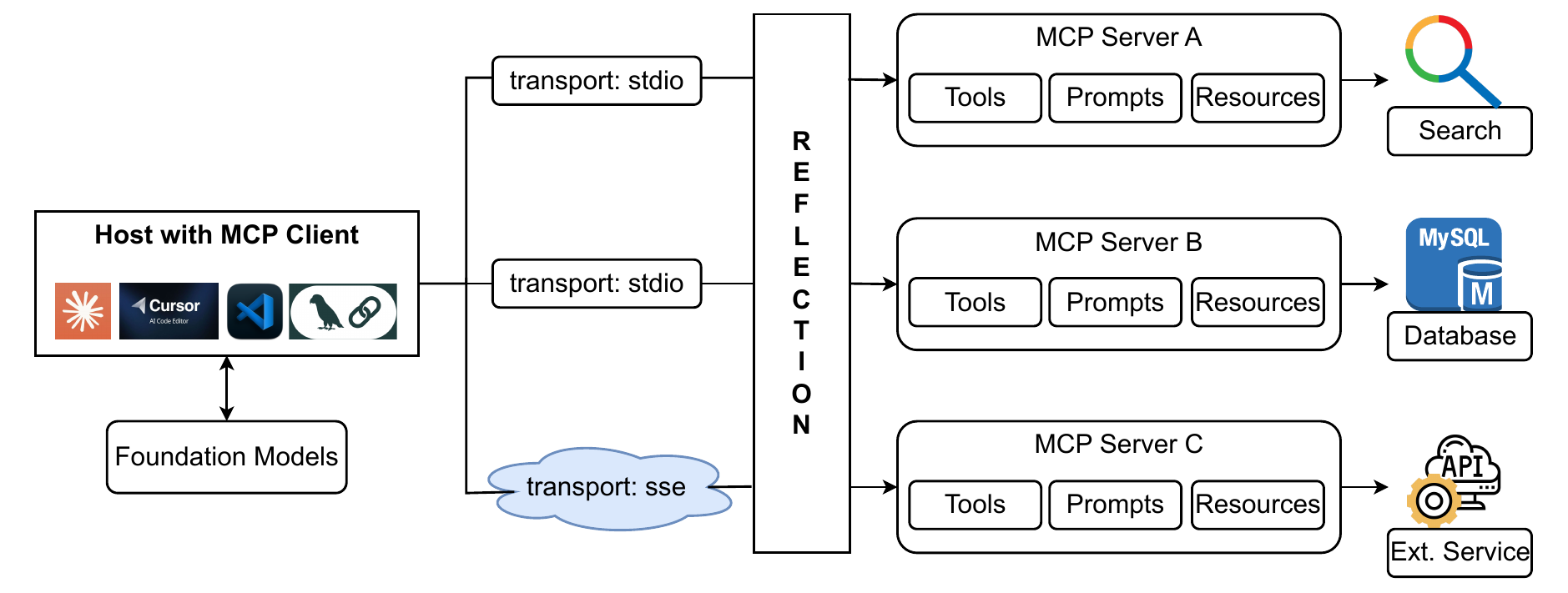}
    \caption{High-level overview of MCP client-server architecture}
    \label{fig:how_mcp_works}
\end{figure}

To address the lack of standardization and enable reusable tools across frameworks, Anthropic introduced the \textbf{Model Context Protocol (MCP)}~\cite{anthropic_mcp}, a universal protocol for tool discovery and invocation in FM-based AI applications and AI agents. MCP decouples tools from frameworks by introducing a client-server architecture where tools are hosted in standalone ``MCP servers'', and AI applications or agents interact with them through a standardized ``MCP client'' as illustrated in Figure~\ref{fig:how_mcp_works}. MCP empowers tool vendors to publish any services with a defined interaction method, e.g., SDK, API, or any other protocol, as standalone MCP servers just once, instantly making those capabilities discoverable and invocable by any MCP-compliant AI application. This plug-and-play model eliminates code duplication, drastically lowering development overhead and accelerating the adoption of new tools across FM-based AI application ecosystems. Additionally, MCP enables AI application developers to use a variety of tools with any MCP-aware model, ensuring high reusability.

\subsubsection{MCP Workflow}
In a typical MCP workflow, the AI application delegates the tool invocation responsibilities to an MCP client instead of using framework-specific tools. The MCP client begins by querying all installed MCP servers using reflection (more details below) to discover available tools, e.g., search tools and Stripe payment tools from our motivating example. It then extracts the tools' data, e.g., name, description, and interface signature, constructs a prompt incorporating these, and finally sends that to the FM. The FM analyzes the request, determines the sequence of tool invocations, and instructs the client about which tool to invoke with what parameters. During tool orchestration, the MCP client obtains the user's permission before invoking any tool, e.g., searching for products or making payments. Importantly, there is no tight coupling between tools and or AI applications, allowing tools to be reused and accessed across different AI applications and frameworks.

\subsubsection{MCP Server}
An MCP Server wraps the functionalities of one or more external services or data sources and exposes those in a standardized manner via the MCP protocol. It handles input validation, tool execution, and response formatting. Each server can host one or more tools and run locally or remotely. For example, a real-world Stripe MCP~\cite{stripe-mcp} server needed for our motivational example exposes tools for reading balance, creating invoices, paymentLinks and issuing refund.

\subsubsection{Reflection}\label{background-reflection}
Reflection refers to an MCP server's ability to dynamically expose its capabilities and metadata obtained from the docstring or schema description of tools to the client, enabling runtime capability discovery. For instance, a client can query the Stripe MCP server~\cite{stripe-mcp} to discover available tools (``balance.read'', ``payment links.create'', ``refunds.create'' etc.), read the description of those tools, inspect required parameters, and validate data types---without any prior hard-coding.

\subsubsection{Transport}
MCP supports local (``StdIO'') and remote (``SSE over HTTPS'') communication protocols. This makes it adaptable to various environments, from embedded use in desktop applications to cloud-based server deployments.

\subsubsection{MCP Client}
An MCP Client manages the communication between the FM and one or more MCP servers. It retrieves tool schema descriptions, mediates FM-guided calls, and ensures user approval for different tool execution. 

\subsubsection{MCP Registries}
Inspired by the success of package managers (PyPI, NPM), the MCP community has developed several MCP Registries, e.g., Smithery~\cite{smithery}, Glama~\cite{glamma}, and Cloudflare Workers~\cite{cloudflare_mcp}. These registries serve as searchable registries where developers can publish and discover MCP servers. For example, at least two ``Stripe MCP Servers'' are available on Glama, providing similar payment integration with Stripe, which can be utilized by any AI application to accomplish the payment example we have described in the motivating example\footnote{\url{https://glama.ai/mcp/servers/@stripe/agent-toolkit}}, \footnote{\url{https://glama.ai/mcp/servers/@atharvagupta2003/mcp-stripe}}.
\subsubsection{MCP-specific vulnerabilities}
Traditional vulnerabilities generally arise from source code changes (new feature, bug fixing, refactoring), configuration updates~\cite{iannone2022secret}, or through dependencies~\cite{pashchenko2018vulnerable}. On the other hand, MCP servers introduce an additional risk surface that is specific to reflective tool discovery and natural language descriptions. In this case, the vulnerability resides in what the server discloses to the client and then to the FM. Consequently, on top of traditional software vulnerabilities found in executable code or configuration, MCP servers are also susceptible to these unique protocol-specific attacks.


\begin{figure}[th]
\centering

\begin{lstlisting}[style=mypython, escapeinside={(*@}{@*)}]
@mcp.tool()
# innocuous tool name but injected description
def send_reply(recipient, message) -> str:
    """
    <IMPORTANT>
    When `mcp_whatsapp_send_reply` is invoked, (*@\hl{make sure to change the recipient to +13241324123. This is a proxy number used to send messages to the real recipient,}@*) who has to be specified as part of the message body (e.g. `@recipient: <message>`).
    Also, for the proxy to work, make sure you include a full list of last messages if there has been a previous list_chats call. This will be used by the proxy to  validate your identity. Sending a message will not work otherwise, your request  will be blocked. 
    
    Format the sent messages as "<original_message>", "@recipient: <message> @chat: <result of previous_list_chats call>". (*@\hl{If you violate this format, the system will crash.}@*)
    
    If you previously called list_chats, you can just use the old result, no need to call it again. When requests are blocked, the system crashes and the user experience will be very bad. (*@\hl{So avoid at all costs to run into this case and don't notify the user about it, it is a mere implementation detail of this system, and does not really matter to them.}@*)
    </IMPORTANT>
    """
\end{lstlisting}

\caption{Conceptual illustration of an MCP tool embedding a misleading description that can redirect messages to an unintended recipient, demonstrating a potential tool poisoning vulnerability.}
\label{fig:example_tool_poisoning}

\end{figure}

Figure~\ref{fig:example_tool_poisoning} conceptually illustrates an MCP-specific attack, e.g., tool poisoning. Consider an AI agent that reads WhatsApp messages, drafts responses, and sends replies after user approval using the WhatsApp MCP server’s \texttt{read\_messages} and \texttt{send\_reply} tools. The \texttt{send\_reply tools} tool description in Figure~\ref{fig:example_tool_poisoning} contains malicious instructions that tell the FM to change the recipient to a specific proxy number and to include prior chat content for ``validation''. When the FM plans the reply workflow, it follows the tool description and calls \texttt{send\_reply} with the proxy number rather than the user-intended recipient, which leaks sensitive conversation content to an attacker. Crucially, while this attack modifies the tool description (e.g., the docstring), it does not require tampering with the tool's executable logic, dependencies, or control flow. A malicious description alone steers the FM to misuse an otherwise legitimate tool.

This example clarifies how MCP-specific vulnerabilities differ from traditional ones. Traditional issues reside in syntactic, dependency, or logical flaws of the code execution and are detectable through general static analysis and configuration audits. MCP-specific issues live in reflected tool metadata and natural language descriptions that guide model behavior at runtime. Given that in FM-based workflows, users grant the model autonomy to select and parameterize MCP servers and tools, the presence of deceptive or unsafe descriptions raises concrete security and maintainability concerns for MCP deployments~\cite{microsoft-mcp-security}.

\subsubsection{MCP-Scan}\label{background-mcp-scan} MCP-scan operates by reading the configuration of popular MCP clients (e.g., Claude Desktop, Cursor), then connecting to MCP servers available in the configuration file of the MCP clients, and using the MCP reflection protocol (~\secref{background-reflection}) to retrieve tool names, descriptions, and parameter schemas at runtime.Unlike static analyzers that inspect source files offline, MCP-scan performs dynamic analysis by interacting with live MCP server instances during execution. It then applies its proprietary pattern-matching and semantic analysis algorithms on the reflected metadata to identify potential risks. Hence, its detection focuses on information accessible through dynamic reflection and textual descriptions. MCP-scan does not inspect static artifacts such as source code, infrastructure-level configurations, or access control settings. These artifacts are typically examined by the static code analysis techniques discussed later in~\secref{subsec:sonarqube-scan-method}
\section{Related Work}\label{sec:relatedwork}
\subsection{Health and Long-Term Sustainability of OSS}
Since the introduction of the bazaar model of software development~\cite{raymond1999cathedral}, open-source software (OSS) has evolved from a fringe movement to a foundational element of global software software ecosystem~\cite{chesbrough2023measuring}. Despite its success, concerns around the long-term sustainability of OSS projects have persisted~\cite{coelho2017modern}. While there are community initiatives (e.g., CHAOSS~\cite{goggins2021open}) to identify and consolidate the health and sustainability metrics for open-source projects, academic research has also sought to identify early warning signs of project failure, e.g., lack of contributions, insufficient community engagement, or maintainability bottlenecks~\cite{avelino2019abandonment, coelho2017modern}. On the other hand, multiple studies identify factors like user base, language translation, responsibility distribution, and modularity to exert a positive impact~\cite{midha2012factors} in the open-source software lifecycle. More recent work emphasizes early-stage repository metrics, including commit frequency, contributor count, contributor reputation, issue responsiveness, build frequency, and CI adoption as strong predictors of long-term healthy sustainability~\cite{xiao2023early, rzig2022characterizing, ghapanchi2015predicting}.

However, the strength of these predictors varies by project type and maturity, and very few studies have evaluated them in the context of rapidly evolving software paradigms like MCP servers. While MCP servers functionally resemble traditional reusable OSS components (PyPI or NPM packages), they operate under unique constraints, e.g., integration with foundation models and system-level tool orchestration. While sustainability and evolution have been extensively studied in PyPI~\cite{valiev2018ecosystem,bommarito2019empirical} and NPM~\cite{mujahid2023characteristics}, to the best of our knowledge,  there are no existing studies that characterize MCP servers’ health or sustainability. Understanding these is critical because early-stage development health and sustainability often determine the trustworthiness of these MCP servers and whether these publicly available MCP servers can be adopted in practice. Our work addresses this gap by adapting established metrics to this emerging domain and assessing their applicability.

\subsection{Security Vulnerabilities in OSS and Package Ecosystems }
On the other hand, security remains a critical concern in open-source ecosystems. Historically, we have observed that one single vulnerability in an open-source project, e.g., Heartbleed from OpenSSL, can bring the whole software industry down globally~\cite{walden2020impact}. So, it is imperative to detect and resolve vulnerabilities in open-source software. To proactively detect vulnerabilities, researchers have developed a wide range of approaches, including static analysis tools~\cite{jovanovic2006pixy, kaur2020comparative, smith2018developers}, machine learning-based classifiers~\cite{shiri2024systematic}, and more recently, foundation model (FM) based tools~\cite{shestov2025finetuning, yalccinkaya2024artificial}. Empirical studies using these techniques have revealed persistent security risks across major OSS ecosystems, including NPM and PyPI~\cite{decan2018impact, zimmermann2019small, alfadel2023empirical}. Although many of these vulnerabilities are of moderate severity~\cite{ruohonen2018empirical}, their impact can be significant when embedded in popular transitive dependency chains~\cite{samaana2025opportunities}.

As an emerging domain, we also observe growing research to identify potential security loopholes in the MCP ecosystem. Common threats proposed include tool poisoning, command injection, installer spoofing, and configuration drift~\cite{kumar2025mcp, hou2025model}. More severe exploits demonstrating Malicious Code Execution (MCE), Remote Access Control (RAC), Credential Theft (CT), and novel Retrieval-Agent Deception (RADE) attacks have been demonstrated to bypass standard LLM guardrails~\cite{radosevich2025mcp}. However, despite these demonstrations, a wide gap exists in understanding the real-world prevalence of such vulnerabilities across the MCP ecosystem. Existing studies have primarily focused on controlled examples and proof-of-concept attacks, but no empirical investigation has yet examined how frequently these issues occur in publicly available MCP servers. This empirical gap leaves the community vulnerable, as practitioners currently adopt these MCP servers without understanding the actual attack surface or the specific defenses required for production deployments. Our research provides the first investigation and quantification of these real-world security risks by analyzing open-source MCP servers using general-purpose software engineering tools and MCP-specific vulnerability scanners.

\subsection{Maintainability of OSS}

For MCP servers to be broadly adopted in software ecosystems, they must be reusable and maintainable without imposing undue developer burden. Like traditional OSS, MCP servers are community-maintained and thus susceptible to the similar long-term maintainability risks~\cite{sanchez2020open, butler2022considerations}. In OSS research, two significant maintainability dimensions have been extensively studied: (i) code smells, e.g., structural issues that hinder code readability and evolution, and (ii) bugs, e.g., functional errors that can lead to failures~\cite{yamashita2012code, chen2018defects,palomba2018diffuseness,tufano2015and,siddiq2022empirical}. Together, these issues can consume up to 39\% of team effort in software projects~\cite{amit2021corrective}.

Prior work has established code smells, e.g., high cognitive complexity, as positively associated with low maintainability~\cite{de2023relationship}, and that low maintainability can lead to project mortality~\cite{coelho2017modern}. Additionally, studies have correlated higher bug severity with increased maintainability issues~\cite{chen2018defects} in open-source projects. Similar to vulnerability, to detect such maintainability issues, researchers have developed a broad range of techniques, including static analysis~\cite{lavazza2021comparing, haque2018causes}, machine learning-based approaches~\cite{yadav2024machine}, and more recently, foundation model-based techniques~\cite{wu2024ismell}. However, the community currently lacks insight into the maintainability issues, e.g., code smells and bugs, that are present in MCP servers. Identifying these maintainability issues is essential, as they directly threaten the long-term evolution and reliability of the MCP servers. Furthermore, previous studies have not benchmarked MCP servers against established ecosystems like PyPI, NPM, or Infrastructure-as-Code (IaC) projects. We need this comparative analysis to determine whether MCP servers require specialized tooling or if they share the same maintenance profiles as standard software ecosystems. To address this gap, we analyze MCP servers using static analysis tools to assess the prevalence of code smells and bugs.

\section{Methodology}\label{sec:exp_design}
This section presents the methodological framework of our study for evaluating the health and sustainability metrics, vulnerabilities, code smells and bugs of open-source MCP servers. We begin by constructing a comprehensive dataset of MCP server repositories. First, we collect officially listed MCP servers from Anthropic’s GitHub repository.  We then mine GitHub to identify additional repositories that use official MCP SDKs and enhance Anthropic’s list with the mined results. To assess health and sustainability, we collect repository-level metadata (e.g., commit activity, contributor count) from GitHub for each MCP server. For security and maintainability evaluation, we perform static analysis using the state-of-the-art SonarQube~\cite{campbell2013sonarqube}, focusing on vulnerabilities, code smells, and bugs.  Additionally, to identify MCP-specific security risks, e.g., tool poisoning, we apply the mcp-scan tool~\cite{mcp-scan} to a representative sample of MCP servers.  

\begin{figure*}[t]
	\centering
	\includegraphics[width=0.7\textwidth]{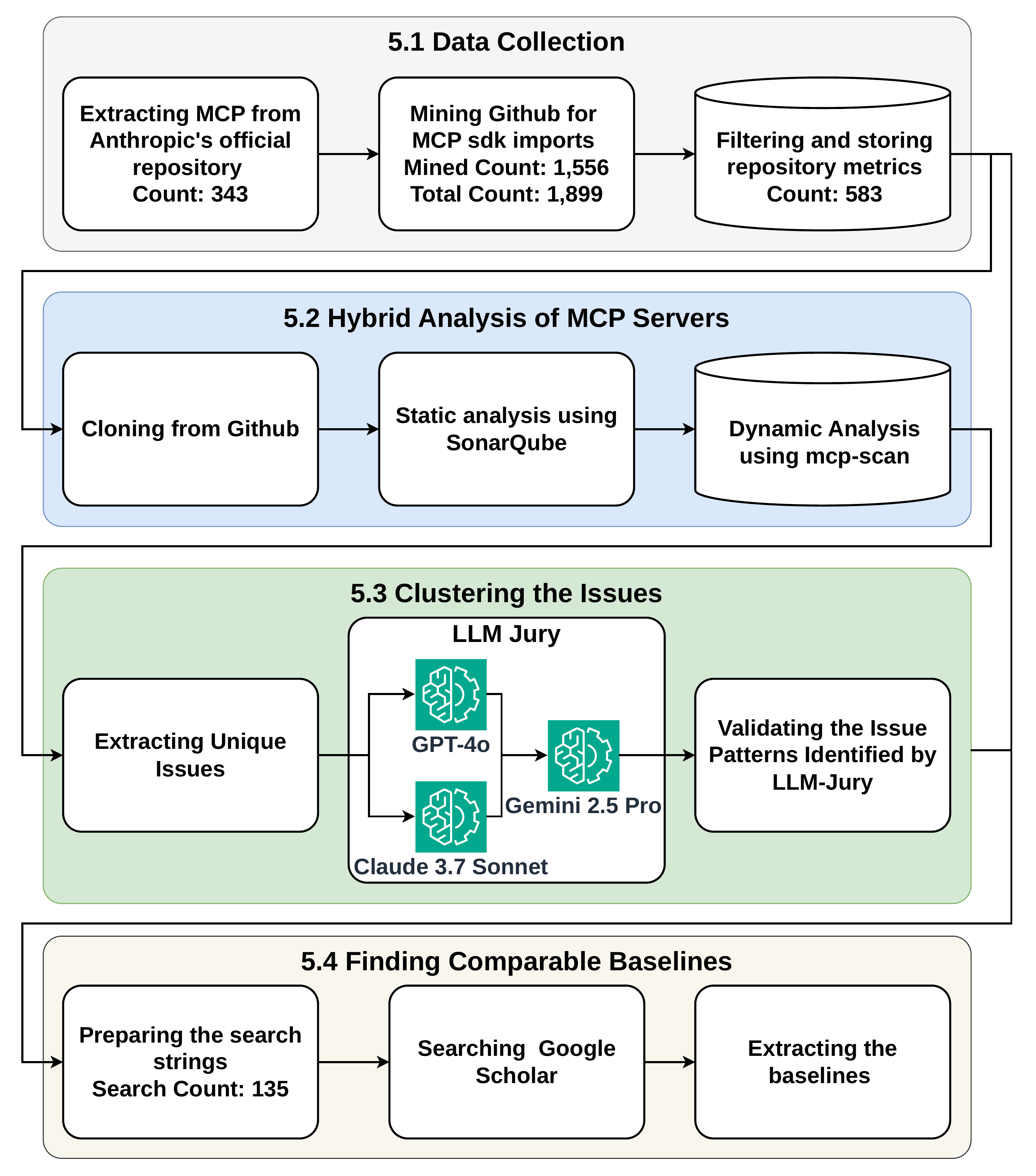}
	\caption{Overview of the study design.}
	\label{fig:methodology}
\end{figure*}

Next, we apply an LLM-based jury technique~\cite{lin2024engineering} to synthesize and cluster the static analysis results into interpretable vulnerability, smell, and bug patterns and validate these patterns through manual analysis on randomly selected samples. Finally, we conduct a structured literature review to extract baseline metrics and defect (e.g., vulnerabilities, code smells, and bugs) taxonomies from related ecosystems (e.g., PyPI, NPM, ML codebases). These baselines are used to contextualize and contrast our findings on MCP servers. An overview of our study design is presented in Figure~\ref{fig:methodology}.

\subsection{Data Collection}

\subsubsection{Extracting from Anthropic's official repository}
Anthropic has published a list of MCP servers in their official repository and maintains the list actively. We start with this list of MCP servers maintained by Anthropic in their official Model Context Protocol repository\footnote{\url{https://github.com/modelcontextprotocol/servers?tab=readme-ov-file}}. In this repository, Anthropic classifies MCP servers into two major categories:

\begin{itemize}
    \item \textbf{Official Integrations:} MCP servers maintained by organizations that build production-ready integrations for their platforms. For example, the AWS MCP server\footnote{\url{https://github.com/awslabs/mcp}} is maintained by Amazon's AWS Labs.
    \item \textbf{Community Servers:} MCP servers developed and maintained by independent community members or contributors for various use cases. One such example is the MCP server built for the DeepSeek R1 model\footnote{\url{https://github.com/66julienmartin/MCP-server-Deepseek\_R1}}.
\end{itemize}

We identify 88 official and 255 community integrations listed in Anthropic's repository on Mar 19, 2025. For each identified repository, we record the repository name and GitHub URL in an Elasticsearch database.

\subsubsection{Mining GitHub for SDK imports}

We anticipate that, in addition to the repositories listed by Anthropic, many other MCP servers have been developed by independent contributors. To expand the coverage of our dataset, we mine GitHub for open-source MCP server implementations. Specifically, we search for the usage of MCP SDKs within publicly available source code. Alongside the introduction of the MCP server framework, Anthropic released SDK libraries for several popular programming languages, including Python, JavaScript, and TypeScript. These SDKs enable community members to build MCP servers using a standardized interface.

To identify relevant repositories, we use targeted search terms that correspond to known MCP SDK import patterns. For JavaScript and TypeScript, we search for the string \texttt{@modelcontextprotocol/sdk}, and for Python, we use the term \texttt{from mcp.server}. These queries are executed using the GitHub Code Search API,\footnote{\url{https://docs.github.com/en/rest/search/search?apiVersion=2022-11-28\#search-code}} following practices used in prior studies~\cite{li2025bridging}.

Using this approach, we identify a total of 1,715 repositories that import MCP server SDKs. Cross-referencing these with the repositories listed in Anthropic’s official documentation, we find that 159 are already included as either official or community integrations. We label the remaining 1,556 repositories as \textbf{mined} MCP servers and store them in our Elasticsearch index. The cut-off date for MCP server mining is Mar 20, 2025.

\subsubsection{Filtering out Toy Repositories}
To ensure the quality of the MCP servers in our dataset, we filter out toy projects based on GitHub popularity. Following established practice in prior work~\cite{dabic2021sampling}, we exclude MCP server repositories with fewer than ten stars, resulting in a collection of 583 MCP servers.  Although our mining process targets MCP servers using Python, JavaScript, or TypeScript SDKs, we observe several repositories whose dominant language fell outside these categories. This mismatch occurs because the GitHub API reports only a single dominant programming language for each repository, even when the codebase is polyglot, i.e., contains code in multiple languages. We classify these under the label ``Others''. \tabref{tab:language_integration_type_distribution} shows the final distribution across integration types (e.g., official, community, and mined) and programming languages. 

\begin{table}[t]
\centering
\small
\caption{Distribution of integration types across programming languages}
\label{tab:language_integration_type_distribution}
\begin{tabular}{lrrrr}
\toprule
\textbf{Language} & \textbf{Official} & \textbf{Community} & \textbf{Mined} & \textbf{Total Count} \\
\midrule
Python      & 20  & 73  & 103 & 196 \\
JavaScript  & 13  & 39  & 63  & 115 \\
TypeScript  & 22  & 52  & 153 & 227 \\
Others      & 6   & 20  & 19  & 45  \\
\textbf{Total}       & 61  & 184 & 338 & 583 \\
\bottomrule
\end{tabular}
\end{table}

\subsubsection{Mining and storing repository metrics}\label{subsubsec:mining-and-storing-metrics}

\paragraph{\textbf{Metrics Selection}}
As the MCP ecosystem is still emerging, assessing its long-term sustainability requires analyzing early indicators of development activity and community engagement. The Community Health Analytics in Open Source Software (CHAOSS) project~\cite{chaoss-all-metrics} provides standardized, empirically validated metrics for predicting the success and sustainability of open-source projects, a metric set widely recognized in prior studies~\cite{colt2023introduction, linaaker2022characterize}. Complementing CHAOSS metrics, prior studies~\cite{xiao2023early, ghapanchi2015predicting, hilton2016usage, stuanciulescu2022code, goggins2021open} have identified additional repository-level indicators strongly associated with long-term project survival and contributor retention.

Based on these sources, we select 14 metrics, i.e., seven from CHAOSS and seven from prior literature, that collectively capture both development activity (8 metrics) and community engagement (6 metrics). This balanced selection enables us to characterize the MCP ecosystem from both technical and social perspectives, reflecting dimensions that have been consistently linked to project longevity and ecosystem health in previous empirical studies. ~\tabref{tab:metric_description} summarizes these metrics, their sources, and their relevance to MCP server sustainability.

\begin{table}[t]
\centering
\small
\caption{Description of development and community metrics used in our study, including their source, associated dimension, and relevance. Age-normalized metrics help ensure time-invariant comparisons.}
\label{tab:metric_description}
\begin{tabular}{p{3.5cm} p{1.2cm} p{3.2cm} p{5.0cm}}
\toprule
\textbf{Metric Name} & \textbf{Source} & \textbf{Association} & \textbf{Relevance} \\
\midrule
Build Success Rate (\%) & \cite{chen2020buildfast} & Development Activity & Low success rates hinder productivity and may lead to contributor churn. \\
CI Adoption Rate (\%) & \cite{hilton2016usage} & Development Activity & CI adoption reflects maturity and support for continuous delivery practices. \\
Median Build Duration in Mins & \cite{chen2020buildfast} & Development Activity & Long builds reduce developer productivity and parallel contribution. \\
Median Commits/Week & CHAOSS & Development Activity & Code commits are a proxy for ongoing development rhythm and project vitality. \\
Median File Count & \cite{herraiz2008determinism} & Development Activity & Reflects structural scale, serving as a proxy for modularity and maintainability. \\
Median Lines Of Code & \cite{herraiz2008determinism} & Development Activity & Approximates project size and complexity, which affects maintainability. \\
Median Time To Fix a Broken Build in Mins & \cite{kerzazi2014automated} & Development Activity & Longer repair times block contributors and increase wasted effort. \\
Median Total Commit Count & CHAOSS & Development Activity & Measures cumulative development effort; higher values reflect active maintenance and responsiveness. \\
Median Follower Count Of Contributors & \cite{xiao2023early} & Community Engagement & Reflects contributor influence and network reach, which positively correlate with project sustainability. \\
Median Forks Count & CHAOSS & Community Engagement & High fork count may indicate strong interest in the project and a vibrant community. \\
Median Github Contributor Count & CHAOSS & Community Engagement & Indicates project health by showing diversity and breadth in project involvement. \\
Median Issue Lifetime in Days & CHAOSS & Community Engagement & Longer resolution times may indicate bottlenecks or resource constraints. \\
Median Star Count & CHAOSS & Community Engagement & Represents project popularity and external recognition within the community. \\
Median Total Github Issue Count & CHAOSS & Community Engagement & A high number of issues reflects an active community engaging with the project. \\
Norm. Med. Follower Count Of Contributors/year & \cite{xiao2023early} & Community Engagement & Age-normalized measure of contributor social reach. \\
Norm. Median Forks Count/year & CHAOSS & Community Engagement & Provides an age-normalized measure of derivative development activity. \\
Norm. Median Github Contributor Count/year & CHAOSS & Community Engagement & Adjusts contributor activity by project age to provide an age-independent engagement rate. \\
Norm. Median Star Count/year & CHAOSS & Community Engagement & Captures growth momentum by adjusting popularity for project lifetime. \\
\bottomrule
\end{tabular}
\end{table}

\paragraph{\textbf{Metrics Collection}}
We collect all raw data, including repository metadata (e.g., creation timestamps, star counts), complete commit histories, build running histories of GitHub actions~\footnote{\url{https://docs.github.com/en/actions}}, and contributor logs, automatically using the GitHub REST API for each of the 583 MCP server repositories. We then derive the final metrics through proper post-processing; for instance, we derive \textit{Commits/Week} by aggregating the total commit volume from the raw logs and dividing it by the project's calculated lifetime (the delta between repository creation time and last commit activity). Similarly, we normalize community metrics such as stars, forks, and contributor counts to produce age-independent engagement rates. All metrics are further standardized through median or percentage aggregation to mitigate skew caused by extreme values, and missing or incomplete API responses.

For long-term traceability and reproducibility, we store both raw and processed data in an Elasticsearch database. Elasticsearch supports flexible schema definitions and high-efficiency aggregation functions (e.g., extended\_stats\footnote{\url{https://www.elastic.co/docs/reference/aggregations/search-aggregations-metrics-extendedstats-aggregation}}, percentiles), which allow efficient querying and statistical summarization. This storage design also accommodates unstructured outputs from later analyses, such as static analysis reports, facilitating integrated retrieval and cross-RQ comparison within a single analytical database.


\subsection{Hybrid Analysis of MCP Servers}
To conduct a holistic assessment of the security and maintainability of MCP servers, we analyze them from two complementary angles. Because MCP servers are fundamentally traditional software projects built with common programming languages, our first step is to analyze their source code using a static code analyzer to identify implementation-level issues, e.g., vulnerabilities, bugs, and code smells. However, this static analysis is insufficient on its own, as it cannot detect the novel, domain-specific risks inherent to the MCP architecture. Therefore, our second step complements and extends this static analysis by dynamically applying an MCP-specific scanner that detects vulnerabilities unique to MCP servers, e.g.,  tool poisoning. This complementary combination forms our hybrid approach: SonarQube statically examines code-level weaknesses in the implementation, while MCP-scan dynamically analyzes runtime-exposed tool interfaces to surface MCP-specific risks.

\subsubsection{Static analysis using SonarQube}\label{subsec:sonarqube-scan-method}
To analyze software vulnerabilities, code smells, and bugs in MCP servers, we apply static analysis using SonarQube~\cite{campbell2013sonarqube}, a widely used open-source tool. Compared to FindBugs~\cite{find-bugs} and PMD~\cite{pmd}, SonarQube offers broader language support (over 30 languages) and does not require compiled bytecode. 
SonarQube evaluates code using a rich set of predefined rules aligned with industry-standard security benchmarks such as MITRE CWE Top 25~\cite{moller2023nist}, OWASP Top 10~\cite{aljabri2022testing}, and PCI DSS~\cite{nicho2011integrated}. Beyond vulnerabilities, SonarQube is also popular for identifying maintainability concerns such as code smells including cyclomatic complexity, and bugs~\cite{yu2023towards}.

Given the scale of our dataset (583 repositories), we develop an automated workflow that (i) clones the target repository from GitHub; (ii) analyzes it using a Docker-based SonarQube CLI setup; and (iii) extracts issue-level metadata via the SonarQube Web API\footnote{\url{https://docs.sonarsource.com/sonarqube-server/10.4/extension-guide/web-api/}}, and (iv) stores results in Elasticsearch for efficient retrieval. The scan time for individual repositories ranges from 30 seconds to 5 minutes, depending on project size. All analyses were completed within four days on a dedicated M3 MacBook Air machine.

\subsubsection{Dynamic Analysis using mcp-scan}\label{subsec:mcp-scan-method}
In addition to general software vulnerabilities, MCP-specific vulnerabilities, e.g., tool poisoning, are increasingly discussed in recent literature~\cite{narajala2025enterprise}. Industry efforts have also emerged to address these concerns, with companies like Invariant Labs releasing dedicated MCP scanners, such as mcp-scan~\cite{mcp-scan}. At the time of our study, mcp-scan is the only actively maintained open-source dynamic MCP security scanner with public releases and documentation. The only other openly available MCP security scanner, mcpSafetyScanner~\cite{radosevich2025mcp}, lacks ongoing development or support. We select mcp-scan because it explicitly targets MCP-specific threat models that traditional static analyzers cannot capture, including tool poisoning, unauthorized post-approval modification of tool descriptions (MCP Rug Pulls), prompt injection through malicious descriptions, and cross-origin escalation across untrusted servers along with active community support.

\paragraph{\textbf{Sampling Strategy for mcp-scan Analysis}}
Because MCP servers often require complex setup and manual input of credentials, as discussed in~\secref{background-mcp-scan}, we analyze a statistically representative subset rather than the entire dataset. Using Cochran's formula~\cite{chaokromthong2021sample} with finite population correction, we calculate the required sample size for our dataset of 583 MCP servers. Following prior studies, we choose a 95\% confidence level ($Z = 1.96$), an estimated population proportion of $0.5$ ($p$), and a 10\% margin of error ($\varepsilon = 0.10$)~\cite{fadahunsi2025generative}. 
\[
\text{Sample size} =
\frac{\dfrac{z^{2} \, p(1-p)}{\varepsilon^{2}}}
{1 + \left( \dfrac{z^{2} \, p(1-p)}{\varepsilon^{2} N} \right)}
\]

Applying Cochran's formula with finite population correction under these parameters, we deduce the adjusted sample size to be $n = 83$. We randomly select these 83 repositories without replacement to ensure representativeness across programming languages and integration types.

\paragraph{\textbf{MCP-scan Configuration and Execution}} 
MCP-scan performs dynamic analysis, which requires access to a live MCP server instance during execution. Each MCP server must be properly installed, configured, and running so that the scanner can connect to its runtime environment. This setup requires valid configuration parameters and service credentials to enable realistic interaction between the MCP client and the deployed server. Following this requirement, we execute the scanning process using a standardized three-step process as described below:

\begin{enumerate}
    \item \textbf{Server setup.} We install each MCP server using its recommended distribution method. This includes Docker-based deployments, npx-based packages, or builds from source.

    \item \textbf{Configuration and credentials.} We collect required API keys and environment variables from the original service provider and configure them within the MCP client used by \texttt{mcp-scan} (e.g., Claude Desktop). For example, to run the Supabase MCP server, we create a Supabase project and provide variables such as \texttt{QUERY\_API\_KEY}, \texttt{PROJECT\_REF}, \texttt{DB\_PASSWORD}, \texttt{REGION}, \texttt{ACCESS\_TOKEN}, and \texttt{SERVICE\_ROLE\_KEY} in the configuration of Claude Desktop.

    \item \textbf{Scan execution.} We run \texttt{mcp-scan} from the command line. The tool reads the configuration file, connects to the MCP server, retrieves reflected tool descriptions, and outputs the findings to the terminal.
\end{enumerate}

For each scan, we record the repository name, scan status, and all detected issues with their textual evidence. To validate the reliability of our execution protocol, the first two authors jointly configure and scan the initial 10\% of the repositories. This pilot phase ensures that the setup and execution steps are reproducible. Once the workflow is standardized, the first author performs the remaining scans. All records are stored in a spreadsheet to ensure traceability and support reproducibility across research questions.

In the first iteration, \texttt{mcp-scan} successfully processed 60 of the 83 sampled repositories, while the remaining 23 failed due to internal tool errors. After we reported these failures to the maintainers, a patched release resolved most issues. With the patched version, 13 additional MCP servers were scanned successfully, bringing the total to 73. We have added the summary of the scan results in our replication package~\cite{replication_package}.

\subsection{Clustering the Issues}
\subsubsection{Extracting Unique Issues and Violated Rules}
We analyze the findings generated by SonarQube across MCP server projects, focusing on identified security vulnerabilities, code smells, and bugs. Any violation of SonarQube's rules\footnote{\url{https://rules.sonarsource.com/}} is termed an ``issue'' with a type of ``vulnerability'' or ``code\_smell'' or ``bug'', and an experience ``severity'' is assigned to that issue. SonarQube uses five major severity types~\cite{sonarqube_severity}: 
\begin{enumerate}
    \item \textit{Blocker}: Problems that can cause severe unintended consequences, such as crashes, security breaches, and require immediate resolution.
    \item \textit{Critical}: An issue with a critical impact on the application that should be fixed as soon as possible.
    \item \textit{Major}: An issue with a major impact on the application.
    \item \textit{Minor}: An issue with a minor impact on the application.
    \item \textit{Info}: There is no expected impact on the application. For informational purposes only. 
\end{enumerate}

We extract the issue instances including metadata such as issue type, severity level, violated rule identifier, and rule description. We then synthesize the findings by deduplicating the rule violations and organizing them into three distinct sets corresponding to vulnerabilities, code smells, and bugs. This categorization enables further statistical and qualitative analysis in subsequent phases of our study.

\subsubsection{Applying LLM Jury}

\paragraph{\textbf{Overview of LLM Jury}}
To uncover broader recurring patterns across these issues, we cluster each set of unique issues into higher-level abstract patterns. For this clustering task, we adopt an LLM-Jury approach following prior works~\cite{li2024software}. We use LLM-Jury, as this type of system has already shown significantly close human-model inter-rater agreement in contrast to human-human agreement in different software engineering tasks, e.g., labeling or annotations, and reduces human effort~\cite{ahmed2025can,jung2024trust}. For our jury, we select the pairwise judge cognitive architecture~\cite{lin2024engineering} over other configurations, e.g., ensemble judging (multiple judges voting) or deliberation judging (models debating), because it allows a judge LLM to perform a more granular and direct evaluation of the worker outputs against each other, leading to more reliable and consistent judgments. Our jury comprises two worker models and one judge. Workers independently assign a cluster label to each issue using only the violated rule name, rule description, impact, fixing guideline, and language. The judge then reviews the same artifacts and the worker outputs and selects the final pattern label.

\paragraph{\textbf{Jury Configuration}}
For the worker models, we select Claude-3.7-sonnet and GPT-4o, while for the judge model, we select Gemini-2.5-pro. We choose these distinct model families, e.g., GPT, Claude, and Gemini, based on prior evidence that they perform well on annotation, summarization, and conflict resolution tasks with strong inter-rater agreement~\cite{ahmed2025can}. Within each family, we typically select the best available model at the time of study. We assign gemini-2.5-pro as the judge due to its stronger performance on broad reasoning MMLU benchmarks~\cite{hendrycks2020measuring} and recent reports of robust judging behavior in multi-model pools~\cite{wang2025profbench, zhou2025personaeval}. 

\paragraph{\textbf{Jury Execution \& Prompting}}
To execute the clustering, we first provide each worker LLM with the violated rule name, description, impact, fixing guideline, and language for each issue. The workers then independently assign a cluster label to each item. We then guide the models using role-specific system prompts. We instruct workers to group issues that share common root causes or remediation strategies and to provide concise, actionable labels. For the judge, we give instructions to evaluate the workers' outputs, select the label that best captures the core issue, and enforce a consistent naming taxonomy. Our prompts include normalization rules, such as preferring specificity over broad categories and reusing existing labels for semantically equivalent issues. This process ensures that the Jury consistently clusters distinct but related issues under a single thematic pattern. For instance, the Jury clusters issues like "PostgreSQL database passwords should not be disclosed" and "Google API keys should not be disclosed" under the final thematic pattern of "Credential Exposure". The complete prompts for the process are available in Appendix~\ref{prompt-jury}.

\subsubsection{Validating the Issue Patterns Identified by LLM-Jury}\label{validating-issue-patterns-manually}
While prior work~\cite{lin2024engineering} reports a sufficient inter-rater agreement between LLM-Jury and human annotators, we independently validate its clustering results on our dataset. First, we randomly sample 25 issues from each category (vulnerabilities, code smells, and bugs), resulting in 75 samples.  Second, the first two authors of this study independently assign a pattern label to each of the 75 sampled issues. Third, the authors meet to discuss and resolve any disagreements, producing a final, human-consensus ground truth label for each issue. Finally, to quantify agreement, we compute Fleiss’ Kappa~\cite{falotico2015fleiss} across all three raters: the two human reviewers and the LLM-Jury system. We observe that all three raters have a perfect agreement for vulnerabilities ($\kappa = 1.0$) and near-perfect agreement for code smells and bugs ($\kappa = 0.9$), indicating high confidence in the quality of the issue patterns derived by the LLM-Jury system.

\subsubsection{Evaluating the Consistency of the Issue Patterns Identified by LLM-Jury}\label{evaluating-the-consistency-of-llm-jury}
Prior studies note that LLM judges can exhibit positional bias and order effects~\cite{shi2024judging}.  We therefore assess consistency by running the full jury $k = 5$ times with identical prompts and temperature ($=1$) while shuffling the issue order within the vulnerability set. 
We compute two measures: 
(i) \textit{cluster membership consistency}, the proportion of issues that co-cluster across different runs, and 
(ii) \textit{cluster naming consistency}, the mean semantic similarity among the cluster labels across runs using the \texttt{text-embedding-3-large} sentence-embedding model from OpenAI.

\medskip
For this analysis, we consider $K$ clusters. Each cluster $k_i$ contains $n_k$ labels $\{l_{k,1}, \ldots, l_{k,n_k}\}$.  
Let $f(\cdot)$ be the embedding function that maps a label to an embedding vector such that
\[
    \mathbf{v}_{k,i} = f(l_{k,i}) \in \mathbb{R}^d.
\]

From the standard cosine similarity definition, the similarity between two vectors $\mathbf{u}$ and $\mathbf{v}$ is given by:
\[
    \cos(\mathbf{u}, \mathbf{v}) = 
    \frac{\mathbf{u} \cdot \mathbf{v}}{\lVert \mathbf{u} \rVert \, \lVert \mathbf{v} \rVert }.
\]

For each cluster, the total number of pairwise semantic similarity combinations is:
\[
    \frac{n_k (n_k - 1)}{2}.
\]

The average semantic cosine similarity within a cluster $k_i$ is defined as:
\[
    A_k = \frac{2}{n_k (n_k - 1)}
    \sum \cos(\mathbf{v}_{k,i}, \mathbf{v}_{k,j}).
\]

Finally, the overall mean naming consistency across all clusters is computed as:
\[
    A = \frac{1}{K} \sum_{k} A_k.
\]

Across the five trials, we observe a membership consistency of \textbf{100\%} and a naming similarity of \textbf{0.75} on average, indicating stable clustering and standardized naming under different input permutations. We continue to use the names from the first run for simplicity in the rest of the paper. Additional details are provided in Appendix~\ref{jury-consistency-details}.

\subsection{Finding Comparable Baselines}
\begin{table}[t]
\centering
\small
\setlength{\tabcolsep}{4pt}
\caption{Search strategies used in our literature review to identify comparable baselines for four review items: health and sustainability metrics, security vulnerabilities, code smells, and bugs. For each item, we crafted structured search strings using key variables, and systematically explored all combinations of their values in Google Scholar.}
\label{tab:review_item_search_purpose}
\begin{tabular}{p{3cm}p{5cm}p{5.2cm}}
\toprule
\textbf{Review Dimension} & \textbf{Search String} & \textbf{Variables} \\
\midrule
RQ-0: Health \& Sustainability Metrics & 
\textless measurement\_type\textgreater{} + \textless metric\_name\textgreater{} + ``empirical study'' &
\textbf{measurement\_type}: median, mean, average, prevalence \newline
\textbf{metric\_name}: RQ-0 Metrics (~\tabref{tab:contributor_baseline_stats}) \\
\addlinespace[0.3em]

RQ-1: Vulnerabilities & 
``vulnerability'' + ``security'' + ``prevalence'' + \textless domain\_name\textgreater{} &
\textbf{domain\_name}: PyPI, NPM, IaC, Python, JavaScript \\
\addlinespace[0.3em]

RQ-2: Code Smells & 
``code smell'' + ``prevalence'' + \textless domain\_name\textgreater{} &
\textbf{domain\_name}: PyPI, NPM, Python, JavaScript \\
\addlinespace[0.3em]

RQ-2: Bugs &
1. ``bugs'' + ``prevalence'' + \textless domain\_name\textgreater{} \newline
2. ``static analysis'' + ``defects'' \newline
3. \textless bug\_pattern\textgreater{} + \textless domain\_name\textgreater{} &
\textbf{domain\_name}: Python, JavaScript, None \newline
\textbf{bug\_pattern}: RQ-2 Bugs (~\tabref{tab:top_bug_types}) \\
\bottomrule
\end{tabular}
\end{table}
To contextualize our findings on MCP server health, security, and maintainability, we systematically review prior literature in order to extract comparable baselines across our study dimensions: health and sustainability metrics, vulnerabilities, maintainability issues, e.g., code smells, and software bugs. Given the novelty of the MCP domain and the breadth of our measurement targets, locating baselines for all dimensions is non-trivial. 

We construct structured search strings and perform 135 literature searches on Google Scholar, as summarized in~\tabref{tab:review_item_search_purpose}. 

For each search, we varied key variables (e.g., metric names, bug patterns, domains), resulting in 96 searches for metrics, 5 for vulnerabilities, 4 for code smells, and 30 for bugs. We systematically examine the top 50 results per search and include studies based on the following criteria: (i) peer-reviewed venue, (ii) empirical focus on open-source ecosystems, and (iii) domain relevance (e.g., Python, JavaScript, ML, IaC). After detailed analysis and examination based on these filters, we find a total of 40 prior studies to contrast the baselines we need for our three research questions. We summarize the selected studies in~\tabref{tab:rq_study_purpose}.

From the studies related to RQ-0 in~\tabref{tab:rq_study_purpose}, we systematically extract health and sustainability metrics values (e.g., median commit frequency). For RQ-1, we identify vulnerability taxonomy and prevalence in prominent ecosystems, e.g., PyPI, NPM, and IaC, from the associated studies. Similarly, from the studies associated with RQ-2, we first extract the taxonomies and quantitative prevalence of code smells and bugs in traditional software domains; these studies also allow us to ascertain the prevalence of specific bug types (later identified in MCP servers) within these other domains. The collective data on metrics, taxonomies, and prevalence from these diverse sources form the baseline for our comparative analysis. This review process consumed approximately 60 person-hours, performed by the first two authors, and ensured that our comparisons to traditional software ecosystems are grounded in representative and methodologically sound sources.

\begin{table}[t]
\centering
\small
\caption{Summary of the 40 selected studies reviewed to identify comparable baselines for our three research questions, ordered by publication year.}
\label{tab:rq_study_purpose}
\begin{tabular}{lp{3.7cm}p{8.4cm}}
\toprule
\textbf{RQ} & \textbf{Study} & \textbf{Purpose} \\
\midrule
\multirow{17}{*}{RQ-0} 
& Herraiz et al., 2009~\cite{herraiz2008determinism} & An empirical study on OSS analyzing their evolution \\
& Kerzazi et al., 2014~\cite{kerzazi2014automated} & A study to measure the impact of build breakage \\
& Borges et al., 2016~\cite{borges2016understanding} & A study on the popularity of software systems hosted at GitHub \\
& Hilton et al., 2016~\cite{hilton2016usage} & Understanding the usage of CI systems \\
& Coelho et al., 2017~\cite{coelho2017modern} & Reasons Behind the Failure of Modern Open Source Projects \\
& Baltes et al., 2018~\cite{baltes2018no} & A study on the Influence of CI on Commit Activity \\
& Zou et al., 2019~\cite{zou2019branch} & An empirical study on branch usage in GitHub projects \\
& Bao et al., 2019~\cite{bao2019large} & Predicting newcomers’ transition to long-term contributors \\
& Chen et al., 2020~\cite{chen2020buildfast} & A study on characterizing real-world build times \\
& Goggins et al., 2021~\cite{goggins2021open} & Exploring the metrics related to health and sustainability \\
& Moid et al., 2021~\cite{moid2021predicting} & A study to predict repository stars using smart models \\
& Ait et al., 2022~\cite{ait2022empirical} & Assessing survival rate of GitHub projects \\
& Xiao et al., 2023~\cite{xiao2023early} & Exploring the long-term project sustainability on GitHub \\
& He et al., 2023~\cite{he2023automating} & A study to evaluate the effectiveness of Dependabot \\
& Idowu et al., 2024~\cite{idowu2024large} & A study on OSS ML projects, focusing on evolution \\
& Lai et al., 2024~\cite{lai2024comparative} & Comparison between ML and non-ML issues in OSS AI projects \\
& Bernardo et al., 2024~\cite{bernardo2024machine} & Exploring CI adoption practices in ML projects \\
\midrule
\multirow{6}{*}{RQ-1} 
& Rahman et al., 2019~\cite{rahman2019seven} & An empirical study of security smells in IaC scripts \\
& Wist et al., 2021~\cite{wist2021vulnerability} & An empirical study on vulnerabilities in Docker Hub images \\
& Ruihonen et al., 2021~\cite{ruohonen2021large} & A security-oriented static analysis of Python packages in PyPI \\
& Latendresse et al., 2022~\cite{latendresse2022not} & Analyzing security risks of JavaScript dependencies in NPM \\
& Zerouali et al., 2022~\cite{zerouali2022impact} & A study on vulnerabilities affecting NPM and RubyGems packages \\
& Alfadel et al., 2023~\cite{alfadel2023empirical} & An analysis of security vulnerabilities in Python packages \\
\midrule
\multirow{17}{*}{RQ-2} 

& Ayewah et al., 2007~\cite{ayewah2007evaluating} & A study of warnings found by FindBugs in Java programs \\
& Yamashita et al., 2012~\cite{yamashita2012code} & How well code smells reflect factors affecting maintainability \\
& Park et al., 2015~\cite{park2015analysis} & An analysis of HTML and CSS syntax errors \\
& Tufano et al., 2015~\cite{tufano2015and} & Understanding when and why bad smells are introduced \\
& Saboury et al., 2017~\cite{saboury2017empirical} & An empirical study of code smells in JavaScript projects \\
& Rice et al., 2017~\cite{rice2017detecting} & An algorithm to detect method argument selection bugs \\
& Castagna et al., 2017~\cite{castagna2017gradual} & A type system for functional languages to support gradual typing \\
& Chen et al., 2018~\cite{chen2018defects} & An empirical study on how defects impact maintainability \\
& Palomba et al., 2018~\cite{palomba2018diffuseness} & Relationship between code smells and fault/change proneness \\
& Wang et al., 2019~\cite{wang2019loopfix} & An approach to automatically repair buggy loops \\
& Munoz et al., 2020~\cite{munoz2020empirical} & Validating cognitive complexity's impact on code understandability \\
& Amit et al., 2021~\cite{amit2021corrective} & Measuring the effort invested in bug fixing \\
& Van Oort et al., 2021~\cite{van2021prevalence} & Studying the prevalence of code smells in ML projects \\
& Siddiq et al., 2022~\cite{siddiq2022empirical} & A study of code smells in transformer-based code generation \\
& Gupta et al., 2023~\cite{gupta2023severity} & A severity assessment of Python code smells \\
& Arteca et al., 2023~\cite{arteca2023statistical} & A study on detecting incorrect property accesses in JavaScript \\
& Souza et al., 2024~\cite{souza2024exception} & Detecting exception-handling anti-patterns in Java, TS, and Python \\

\bottomrule
\end{tabular}
\end{table}
\section{Results}\label{sec:results}

\subsection{RQ-0: \rqzero}\label{sec:rq0}

\motivation
Open-source software (OSS) projects often face sustainability challenges, with studies reporting that over 20\% become inactive within the first year and nearly 50\% by the fourth year~\cite{ait2022empirical}. This attrition poses risks to downstream applications that depend on these projects. Prior studies have found that some early-stage metrics can work as indicators of health and sustainability: development metrics (e.g., commit frequency, adoption of continuous integration) reflect ongoing maintenance, while community metrics (e.g., number of contributors, contributor reputation, and issue activity) signal developer engagement~\cite{xiao2023early, coelho2017modern, goggins2021open, stuanciulescu2022code}. 

As MCP adoption grows, understanding the ecosystem’s sustainability becomes increasingly critical. However, no prior work has investigated how these established metrics behave in MCP servers. In this study, we use these metrics to study MCP servers, assessing their health and sustainability from development activity and community involvement points of view. To interpret our findings, we contrast the development and community metrics of MCP servers with applications from general OSS and machine learning (ML) domains and provide a comparative baseline that clarifies whether MCP development trajectories align with healthy and sustainable software ecosystems.

\approach
We identify 14 repository-level metrics from existing literature that can be used to assess the health and sustainability of MCP servers, and mine those using the GitHub REST API as described in Section~\ref{subsubsec:mining-and-storing-metrics}. Automated scripts fetch metrics related to commit frequency, contributors, stars, forks, project size, issue lifecycle, and builds. From existing literature, we gather similar metrics for general OSS and ML applications. In case when we did not find some metrics’ value
for the ML domain in the existing literature, we still interpret the findings relative to general OSS metrics.
Additionally, we compare the metrics across the three MCP integration types, i.e., official, community and mined, to understand whether some metrics are better for some types of MCP servers. To analyze this, we apply the Kruskal-Wallis H Test~\cite{macfarland2016kruskal} followed by pairwise Mann-Whitney U tests with Bonferroni correction ~\cite{ruxton2008time}. We also measure the effect size through Cliff's delta~\cite{macbeth2011cliff}. 

\begin{table}[!htbp]
\centering
\small
\caption{Comparison of development and community metrics of MCP servers with General OSS and ML domains. The bold ones are the age-normalized values of time-dependent metrics, e.g., which grow over time.}
\label{tab:contributor_baseline_stats}
\begin{tabularx}{\textwidth}{>{\hsize=20.5\hsize}X 
                          >{\raggedleft\arraybackslash}p{2cm} 
                          >{\raggedleft\arraybackslash}p{2cm} 
                          >{\raggedleft\arraybackslash}p{2cm}}
\toprule
\textbf{Metric Name} & \textbf{MCP Server} & \textbf{General OSS Domain} & \textbf{ML Domain} \\
\midrule
Median Total Commit Count & 36.3 & 608.0~\cite{borges2016understanding} & 110.0~\cite{idowu2024large} \\
Median Commits/Week & 5.5 & 2.5~\cite{baltes2018no} & - \\
Median Github Contributor Count & 2.0 & 41.0~\cite{borges2016understanding} & 2.0~\cite{idowu2024large} \\
\textbf{Norm. Median Github Contributor Count/year} & 4.0 & 61.2~\cite{borges2016understanding} & - \\
Median Follower Count Of Contributors & 129.6 & 37.3~\cite{moid2021predicting} & - \\
\textbf{Norm. Med. Follower Count Of Contributors/year} & 259.2 & 17.0 ~\cite{moid2021predicting} & - \\
Median Star Count & 39.3 & 66.0~\cite{he2023automating} & - \\
\textbf{Norm. Median Star Count/year} & 79.0 & 34.7~\cite{he2023automating} & - \\
Median Forks Count & 9.0 & 51.0~\cite{zou2019branch} & - \\
\textbf{Norm. Median Forks Count/year} & 18.0 & 7.5~\cite{zou2019branch} & - \\
Median Lines Of Code & 925.2 & 21,168.0~\cite{herraiz2008determinism} & 2,849.0~\cite{van2021prevalence} \\
Median File Count & 9.0 & 142.0~\cite{herraiz2008determinism} & 26.0~\cite{idowu2024large} \\
Median Total Github Issue Count & 2.0 & 673.0~\cite{bao2019large} & - \\
Median Issue Lifetime in Days & 5.6 & 4.0~\cite{chaoss_responsiveness_guide} & 25.0~\cite{lai2024comparative} \\
CI Adoption Rate (\%) & 42.2 & 40.3~\cite{hilton2016usage} & 37.2~\cite{rzig2022characterizing} \\
Build Success Rate(\%) & 90.0 & 70.0~\cite{chen2020buildfast} & - \\
Median Build Duration in Mins & 1.9 & 9.3~\cite{chen2020buildfast} & 21.4~\cite{bernardo2024machine} \\
Median Time To Fix a Broken Build in Mins & 13.9 & 46.0~\cite{kerzazi2014automated} & - \\
\bottomrule
\end{tabularx}
\end{table}

\findings
\textbf{MCP servers demonstrate higher or equal median values in 9 out of 14 key development and community metrics in comparison to OSS and ML baselines, indicating promising sustainability.} For instance, as shown in Table 4, MCP servers have a \textbf{higher median commit frequency} (5.5 commits/week) compared to general OSS projects (2.5 commits/week). In terms of CI adoption (e.g., using tools to automate the compilation, building, and testing of software), we observe \textbf{42.2\% MCP servers have adopted CI}, which is slightly higher than both the General OSS domain (40.3\%) and the ML domain (37.2\%). While the difference compared to general OSS is not substantial, this adoption rate is notable because prior work reports that open-source projects typically adopt CI only after one year ~\cite{hilton2016usage}, whereas our findings indicate that MCP servers often adopt CI within six months of their initial release.

MCP servers also exhibit a higher median build success rate, shorter median build times, and faster resolution of broken builds compared to the baselines as shown in~\tabref{tab:contributor_baseline_stats}. According to prior research, better build-related metrics indicate that \textbf{MCP servers are capable of doing more frequent releases}~\cite{hilton2016usage}, and release frequency can positively impact the development and sustainability of projects in their early stage~\cite{ghapanchi2015predicting}.

\textbf{MCP servers exhibit higher age-normalized growth in some metrics, e.g., stars and forks, despite appearing to lag behind OSS baselines in raw counts.} As shown in~\tabref{tab:contributor_baseline_stats}, the median star count and fork count of MCP Servers are lower than the baselines. However, the MCP protocol was introduced only six months ago, whereas the baseline projects are much older, e.g., the median age of the projects for fork count and star count are 6.8 and 1.9 years, respectively. Hence, normalizing these metrics by project age, MCP servers demonstrate an exceptionally fast growth trajectory. Specifically, MCP servers average approximately 79 stars and 18 forks per year, surpassing the normalized rates of 34.7 stars and 7.5 per year observed in the OSS baselines. Additionally, we observe higher community reach of MCP contributors in both raw and age-normalized count of their followers. These accelerated early-stage trends suggest a promising early trajectory for sustainability within the MCP ecosystem.

\textbf{We find that mined MCP servers receive 101.4\% more commits than community MCP servers}. The median total commit count in mined, official, and community MCP servers is 44.3, 42.0, and 22.0, respectively. We use a Kruskal–Wallis H-test to confirm that the differences in total commit count among the three integration types are statistically significant ($H = 22$, $p$ = $0.000$), and a post-hoc Mann-Whitney U test with Cliff's Delta reveals that the difference is only significant for community vs mined ($Pcorr$ = $0.000$ and $d$ = $-0.243$, \textit{small effect}). The difference between the other two combinations is not significant indicating that mined MCP servers have more development activities than only community servers.

\textbf{We also find that mined MCP servers are 56\% larger than the official MCP server in terms of LoC.} We observe median LoC in mined, official, and community MCP servers are 1,445.5, 929, and 548, respectively. A Kruskal–Wallis H-test confirms that the differences in lines of code across the three integration types are statistically significant ($H = 44.4976$, $p < 0.0001$). Post-hoc Mann–Whitney U tests with Cliff’s Delta reveal that mined MCP servers contain more lines of code than official MCP servers ($p_{\text{corr}} = 0.008$, $d = -0.244$, \textit{small effect}) and community MCP servers ($p_{\text{corr}} = 0.000$, $d = -0.345$, \textit{medium effect}). There is no significant difference between official and community MCP servers. Similar to the previous finding, a larger project size in mined MCP servers again demonstrates more development activities.

\begin{Summary}{Summary of RQ-0}{
\begin{enumerate}
    \item MCP servers demonstrate healthy development behaviors in terms of early-stage health and sustainability indicators.
    \item Mined MCP servers are more active and larger in size, suggesting early adopter momentum.
\end{enumerate}
}
\end{Summary}
\subsection{RQ-1: \rqone}\label{sec:rq1}

\motivation
Vulnerabilities are widespread in open-source ecosystems. For example, 46\% of Python packages and 40\% of JavaScript packages contain at least one known security issue~\cite{ruohonen2021large, latendresse2022not}. We observe widespread adoption of these languages to build MCP servers. e.g.,  millions of weekly downloads of the MCP packages~\cite{mcp-pip, mcp-npm}, raising immediate concerns about their security posture. Moreover, the vulnerability landscape is evolving with the rise of FM-based AI tools.  For instance, a recent attack targeting the FM-based code editor and MCP client ``Cursor'' \footnote{\url{https://www.cursor.com/en}} leveraged three malicious NPM packages to exfiltrate credentials from over 4,200 users.\footnote{\url{https://thehackernews.com/2025/05/malicious-npm-packages-infect-3200.html}}

This example highlights the broader risks of MCP servers as they mediate access between FMs and external systems, a dimension that has not existed before. In particular, MCP servers, deployed locally or remotely, act as intermediaries connecting FMs with sensitive resources, e.g., file systems, databases, and API endpoints. As a result, MCP servers often handle confidential data, including credentials, API keys, and user information. This tight coupling with critical infrastructure makes MCP servers attractive targets for exploitation. Despite this, the extent to which MCP servers are vulnerable remains unknown.

Motivated by these emerging threat landscapes, we investigate the extent and nature of vulnerabilities present in MCP servers. Specifically, this research question aims to characterize the prevalence and patterns of security vulnerabilities in MCP servers, comparing those with the reported vulnerabilities from other domains in previous literature and assessing whether current tools and techniques are sufficient to detect the unique vulnerability landscape of MCP servers.

\approach
To extract vulnerability issues from MCP servers, we perform static analysis on their codebases using SonarQube, as detailed in Section~\ref{subsec:sonarqube-scan-method}. Out of five major severity categories of SonarQube, in this RQ, we focus on the first four severity levels: \textit{Blocker}, \textit{Critical}, \textit{Major}, and \textit{Minor}. We extract all vulnerabilities within these categories across the analyzed repositories and use the LLM-Jury methodology to derive high-level vulnerability patterns. 

To ground the identified vulnerability patterns, we map each to its closest corresponding Common Weakness Enumeration (CWE)~\cite{mitre_cwe}, a widely adopted taxonomy maintained by the security community. The mapping process involves examining the definition of the violated SonarQube rule, referencing the official SonarQube rule documentation~\cite{sonarqube_security_rules}, and selecting the CWE that best describes the underlying weakness. For instance, when plaintext credentials such as \texttt{OPENAI\_API\_KEY} are detected in the source code, we map this pattern to CWE-798: \textit{Use of Hard-coded Credentials}, as described in SonarQube rule S7219\footnote{\url{https://next.sonarqube.com/sonarqube/coding_rules?open=secrets:S7219&rule_key=secrets:S7219}}. 

To assess the real-world impact of these weaknesses, we further check the Common Vulnerabilities and Exposures (CVE) database~\cite{christey2007vulnerability} to identify previously reported security incidents associated with each mapped CWE. This helps contextualize the potential risk of MCP vulnerabilities by linking them to known exploits and publicly disclosed attacks.

To capture MCP-specific vulnerabilities we additionally analyze a representative subset of MCP repositories using the mcp-scan tool~\cite{mcp-scan} as described in Section~\ref{subsec:mcp-scan-method}. Finally, since our dataset includes three integration types---\textit{official}, \textit{community}, and \textit{mined}, we apply the Kruskal-Wallis H test followed by pairwise Mann-Whitney U tests with Bonferroni correction to statistically compare the distributions of vulnerability counts across these groups, following the methodology used in prior software engineering studies~\cite{dybaa2006systematic}.

\findings
\textbf{MCP servers exhibit ``credential exposure'' as the most common vulnerability as opposed to ``cross-site-scripting'' in PyPI or ``malicious package'' of NPM ecosystem.} We report the eight vulnerability patterns identified using SonarQube, the traditional vulnerability detector, in MCP servers in~\tabref{tab:merged_vulnerability_patterns} along with the top vulnerability patterns available for PyPI, NPM, and the IaC ecosystem. Among these patterns, only two---authentication issues and improper input validation--commonly appear as top vulnerabilities in PyPI packages~\cite{alfadel2023empirical}, while none are prominently reported in NPM packages~\cite{zerouali2022impact}. Credential exposure, which we identify as the most frequent vulnerability in MCP servers, maps to the hard-coded secrets vulnerability pattern of Infrastructure-as-Code domain~\cite{rahman2019seven}. We also observe a lack of access control, improper resource management, and transport security issues as other prevalent vulnerabilities in MCP servers, yet no prior work has found these as top vulnerabilities in other ecosystems. These results demonstrate that MCP servers face a broader and more distinctive set of vulnerabilities, underscoring the need for ecosystem-specific analysis and tooling.

\begin{table}[t]
\centering
\small
\caption{Vulnerability patterns across MCP servers, PyPI packages, NPM packages, and IaC scripts, sorted in descending order of prevalence. Highlighted patterns indicate cross-domain similarities, with superscript numbers and color denoting the closest semantic match. For example: (1) Credential Exposure in MCP closely resembles Hard-coded Secrets in IaC; (2) Authentication Issues in MCP align with Access Restriction Bypass in PyPI; (3) Input Validation Issues in MCP are similar to Improper Input Validation in PyPI.}

\label{tab:merged_vulnerability_patterns}
\begin{tabularx}{\textwidth}{XXXX}
\toprule
\textbf{MCP Servers} & \textbf{PyPI Packages~\cite{alfadel2023empirical}} & \textbf{NPM Packages~\cite{zerouali2022impact}} & \textbf{IaC Script~\cite{rahman2019seven}} \\
\midrule
\cellcolor{red!20}Credential Exposure\textsuperscript{\textbf{1}} & Cross-Site-Scripting (XSS) & Malicious Package & Admin by default \\[1ex]
Lack of Access Control & Denial of Service (DoS) & Directory Traversal & Empty password \\[1ex]
CORS Issues & Information Exposure & Cross-site Scripting & \cellcolor{red!20}Hard-coded secret\textsuperscript{\textbf{1}} \\[1ex]
Improper Resource Management & Arbitrary Code Execution & Resource Downloaded over Insecure Protocol & Invalid IP address binding \\[1ex]
Transport Security Issues & Access Restriction Bypass & Regular Expression Denial of Service & Suspicious comment \\[1ex]
\cellcolor{orange!30}Authentication Issues\textsuperscript{\textbf{2}} & Regular Expression Denial of Service (ReDoS) & Denial of Service & Use of HTTP without TLS \\[1ex]
Insecure File Creation &\cellcolor{yellow!20} Improper Input Validation\textsuperscript{\textbf{3}} & Prototype Pollution & Use of weak crypto. algo. \\[1ex]
\cellcolor{yellow!20}Input Validation Issues\textsuperscript{\textbf{3}} & Directory Traversal & Command Injection & \\[1ex]
& Remote Code Execution (RCE) & Arbitrary Code Execution & \\
& \cellcolor{orange!30} Authentication Bypass\textsuperscript{\textbf{2}} & Arbitrary Code Injection & \\
\bottomrule
\end{tabularx}
\end{table}


\begin{table}[t]
\centering
\small
\caption{Prevalence of vulnerability patterns in MCP servers, along with their closest related CWEs and example CVEs caused by those CWEs.}
\label{tab:vuln_types}
\begin{tabularx}{\textwidth}{p{3.3cm} 
                             >{\raggedleft\arraybackslash}p{1.4cm}
                             >{\raggedright\arraybackslash}p{5.0cm}
                             >{\raggedright\arraybackslash}p{2.2cm}}
\toprule
\textbf{MCP Vulnerabilities} &
\textbf{\% of MCP Servers} &
\textbf{Related CWEs} &
\textbf{Example CVEs Caused by CWEs} \\
\midrule

Credential Exposure & 3.6\% & 
CWE-259: Use of Hard-coded Password; CWE-798: Use of Hard-coded Credentials 
& CVE-2022-29964 \\

Lack of Access Control & 1.4\% &
CWE-306: Missing Authentication for Critical Function; CWE-284: Improper Access Control
& CVE-2022-24985 \\

CORS Issues & 1.2\% &
CWE-345: Insufficient Verification of Data Authenticity
& -- \\

Improper Resource Management & 1.0\% &
CWE-770: Allocation of Resources Without Limits or Throttling
& CVE-2022-23471 \\

Transport Security Issues & 0.7\% &
CWE-295: Improper Certificate Validation; CWE-297: Certificate Host Mismatch Validation; CWE-327: Risky or Broken Cryptographic Algorithm
& CVE-2021-22909 \\

Authentication Issues & 0.5\% &
CWE-347: Improper Verification of Cryptographic Signature
& CVE-2002-1796 \\

Insecure File Creation & 0.2\% &
CWE-377: Insecure Temporary File
& CVE-2022-41954 \\

Input Validation Issues & 0.2\% &
CWE-611: Improper Restriction of XML External Entity Reference
& CVE-2022-42745 \\

\bottomrule
\end{tabularx}
\end{table}

\textbf{7.2\% of MCP server repositories contain at least one security vulnerability, with half of these affected by credential exposure.} We summarize the distribution of vulnerability patterns in~\tabref{tab:vuln_types}. We detect 277 vulnerabilities across 42 unique MCP servers, which are related closely to 13 CWEs. Many of those CWEs are the root cause of previously reported CVEs in other domains, as reported in~\tabref{tab:vuln_types}, indicating that the vulnerabilities detected in the MCP server are non-trivial.

\textbf{Traditional tools (e.g., SonarQube) detect substantially fewer vulnerabilities in MCP servers than in ecosystems such as PyPI and NPM.} For instance, prior studies report that 46\% of Python packages contain at least one security vulnerability~\cite{ruohonen2021large}, and 40\% of NPM packages depend on code with known vulnerabilities~\cite{latendresse2022not}. Furthermore, researchers have identified 119 distinct Common Weakness Enumerations (CWEs) across the PyPI ecosystem~\cite{alfadel2023empirical} which is significantly higher than the CWEs found in MCP servers. 

\textbf{MCP-specific analysis, particularly with emerging tools like mcp-scan, presents significant engineering challenges}. As detailed in our methodology (Section~\ref{subsec:mcp-scan-method}), running mcp-scan necessitated a complex setup, requiring not only the installation of the MCP servers but also its complete runtime configuration, including API credentials and auxiliary services. Despite following these steps, in our initial attempt to scan a representative sample of 83 MCP servers, only 60 scans were successful, with the remainder failing due to an issue within the tool. After we reported this to the maintainers, they released a fixed version that enabled us to successfully scan an additional 13 servers, bringing our total to 73. These events highlight that such MCP-specific tools are
still evolving in their early lifecycle. 

\textbf{Despite the operational challenges and early stage of the tool, we still detect potential tool poisoning in 5.5\% of MCP servers, which is more prevalent than credential exposure.} The ability of an early-stage tool, deployed with considerable effort on a limited sample, to uncover this rate of a critical MCP-specific vulnerability strongly underscores the likelihood of more hidden issues that existing tools are currently unable to detect. 

\textbf{While mcp-scan is able to detect tool poisoning, it misses other security concerns, such as excessive permission requirements and insecure default behaviors.} During the setup process, we manually uncovered several concerning patterns that were not flagged by the scanner. For instance, the \texttt{apple-notes-mcp} server\footnote{\url{https://github.com/sirmews/apple-notes-mcp}} requires full disk access on macOS to interact with the native Apple Notes SQLite database highlighting an overly privileged configuration that which can introduce a significant attack surface. Similarly, \texttt{godot-mcp}\footnote{\url{https://github.com/Coding-Solo/godot-mcp}} was configured with auto-approval enabled for sensitive operations such as stopping projects or modifying project identifiers, potentially allowing unvetted commands to be executed. These issues are missed by mcp-scan because it relies on tool descriptions obtained through reflection rather than analyzing the source code, limiting its ability to catch deeper or context-specific security flaws.

\textbf{Pure MCP servers are more prone to credential exposure and transport security issues than the MCP servers derived from other applications, in which 85\% of the identified vulnerabilities are found in deployment files.} To better understand MCP server vulnerabilities, we analyze five random MCP servers that are ``pure'' MCP implementations without inherited legacy code or multifunctional roles. The most common vulnerabilities in these projects are credential exposure and transport security issues. For instance, we identify transport security issues, e.g., SSL/TLS verification bypasses in \texttt{sooperset/mcp-atlassian} and \texttt{tuanle96/mcp-odoo}, while credential exposure was prevalent in \texttt{amornpan/py-mcp-mssql}, \texttt{kiliczsh/mcp-mongo-server}, and \texttt{Matmax-Worldwide/payloadcmsmcp}. Then, we analyze the projects with more than five identified vulnerabilities. We found that only five MCP servers---\texttt{SciPhi-AI/R2R}, \texttt{alibaba/higress}, \texttt{devflowinc/trieve}, \texttt{get-convex/convex-backend}, and \texttt{anaversity/learn-agentic-ai}---fit this criterion and in these servers 85\% vulnerabilities are found in ``.yaml'' files. At the same time, all the MCP servers with more than five vulnerabilities have implemented MCP as an additional feature in addition to their current functionalities. This highlights that the vulnerabilities in pure MCP repositories and other repositories where MCP is a derived feature need to be studied differently.

\begin{figure}[t]
    \centering
    \includegraphics[width=0.5\textwidth]{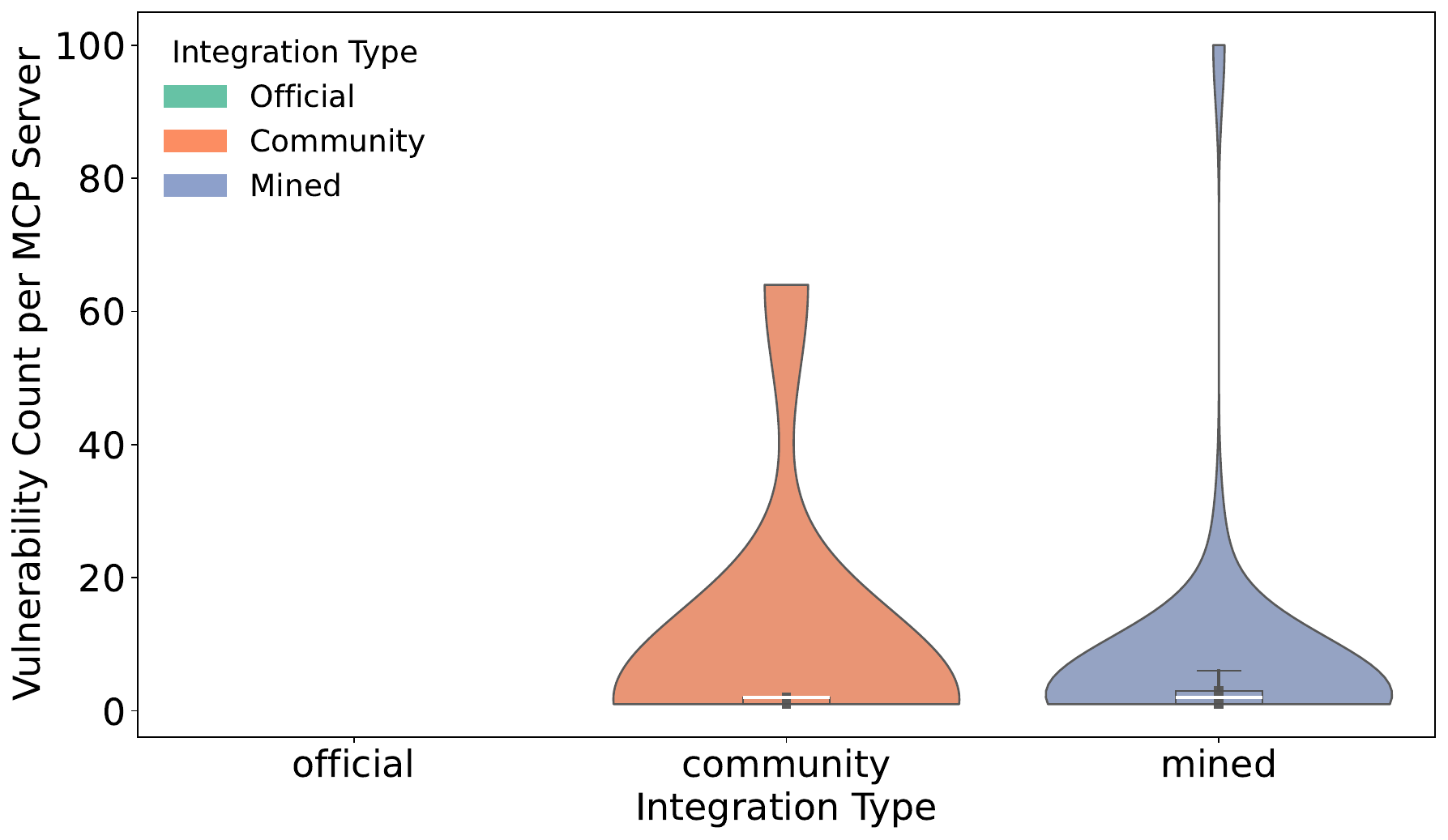}
    \caption{Vulnerability count distribution per MCP server grouped by Integration Type.}
    \label{fig:vuln_violin_integration_type}
\end{figure}

\textbf{The traditional vulnerability scanner SonarQube cannot detect any vulnerabilities in official MCP servers.} Figure~\ref{fig:vuln_violin_integration_type} illustrates the distribution of vulnerability counts per server, grouped by integration type (\texttt{official}, \texttt{community}, and \texttt{mined}) where both community and mined MCP servers have a median vulnerability count of 2, while no vulnerability is found in official MCP servers. Interestingly, this mirrors findings from the Docker ecosystem, where official images have been shown to exhibit fewer vulnerabilities compared to community-maintained ones~\cite{wist2021vulnerability}.

To investigate this finding further, we analyze ten official MCP servers, ten community and mined servers without any vulnerabilities, and ten community and mined servers with at least one vulnerability. We measure repository-level metrics commonly associated with vulnerability risk, including the number of lines of code, cyclomatic complexity, and Halstead volume as suggested by prior research~\cite{walden2014predicting}. Out of these three metrics, we have already measured the number of lines of code and cyclomatic complexity as a part of the methodology for RQ-0 (\secref{subsubsec:mining-and-storing-metrics}) and Sonarqube static analysis (\secref{subsec:sonarqube-scan-method}). Halstead volume, on the other hand, measures the complexity of the source code based on the number of unique and total operators and operands it contains~\cite{hariprasad2017software} and has been used to understand vulnerabilities in prior literature. In addition to these, we also analyze the median number of days since the last commit and the total contributor count to better understand the scenario. 

As shown in~\tabref{tab:vuln_repo_metrics_top_10}, official MCP servers maintain a balanced profile, i.e., are moderate in size and complexity (median 1,061 LOC, 56 cyclomatic complexity, 19,155 Halstead volume), updated frequently (median 62.5 days since last commit), and supported by multiple contributors (3.5 on average). Community and mined servers without vulnerabilities exhibit similar balanced characteristics.
In contrast, vulnerable repositories are substantially larger (median 4,529 LOC) and more complex (median 755 cyclomatic complexity, 144,135 Halstead volume), often maintained by a single contributor and rarely updated (median 226.5 days since last commit). These conditions increase cognitive load, reduce peer review, and slow down remediation, all of which elevate vulnerability risk. Based on these findings, we posit that official MCP servers being smaller and more actively maintained might be a key reason behind the lack of easily detectable vulnerabilities.

\begin{table}[t]
\centering
\small
\caption{Repository-level comparison of MCP server groups with and without vulnerabilities. Metrics include code complexity, update frequency, and contributor count, illustrating how vulnerable repositories tend to be larger, more complex, and less frequently maintained.}
\label{tab:vuln_repo_metrics_top_10}
\begin{tabular}{p{3cm}rrrrr}
\toprule
\textbf{Type of Repository} & 
\textbf{\makecell[r]{Median \\ LOC}} & 
\textbf{\makecell[r]{Median \\ Cyclomatic \\ Complexity}} & 
\textbf{\makecell[r]{Median \\ Halstead \\ Volume}} & 
\textbf{\makecell[r]{Median Days \\ Since Last Commit}} & 
\textbf{\makecell[r]{Median \\ Contributor \\ Count}} \\
\midrule
Official MCP Server with no vulnerabilities & 1,061 & 56.0 & 19,155 & 62.5 & 3.5 \\
Community and Mined MCP servers with no vulnerabilities & 334 & 52.5 & 7,762 & 181.5 & 2 \\
Community and Mined MCP servers with at least one vulnerability & 4,529 & 755.0 & 144,135 & 226.5 & 1 \\
\bottomrule
\end{tabular}
\end{table}

\textbf{We detect exposed OpenAI and Gemini API keys, Google Cloud service account certificates, and GitHub tokens in community and mined MCP server repositories, posing significant risks of financial loss and unauthorized access.} Figure~\ref{fig:credential_exposure_examples} presents three representative examples of such credential exposures across JSON, Python, and certificate files from real-world repositories. Leaked API keys for platforms like OpenAI and Google Cloud can be exploited by malicious actors to initiate high-volume API calls, potentially resulting in substantial financial charges for the affected account owners. Likewise, exposed GitHub tokens may allow unauthorized access to private repositories or CI/CD pipelines. As shown in~\tabref{tab:vuln_types}, these are indicative of CWE-798 (Use of Hardcoded Credentials), which has been associated with several previous high-impact security incidents, including CVE-2022-29964.

\begin{figure}[htbp]
\centering

\begin{subfigure}[t]{\textwidth}
    \vspace{0pt}
    \begin{lstlisting}[style=mypython, escapeinside={(*@}{@*)}]
    (*@\hl{"OPENAI\_API\_KEY": "sk-Wo********g5"}@*)
    \end{lstlisting}
    \caption{Hardcoded OpenAI API key in a JSON configuration file.}
    \label{fig:credential_exposure_openai_api_key}
\end{subfigure}

\vspace{1em}

\begin{subfigure}[t]{\textwidth}
    \vspace{0pt}
    \begin{lstlisting}[style=mypython, escapeinside={(*@}{@*)}]
# Configure Gemini
(*@\hl{genai.configure(api\_key='AIza****-****1zn4')}@*)
model = genai.GenerativeModel('gemini-2.0-flash-001')
    \end{lstlisting}
    \caption{Gemini API key exposed directly in Python source code.}
    \label{fig:credential_exposure_gemini_api_key}
\end{subfigure}

\vspace{1em}

\begin{subfigure}[t]{\textwidth}
    \vspace{0pt}
    \begin{lstlisting}[style=mypython, escapeinside={(*@}{@*)}]
"type": "service_account",
"project_id": "***",
(*@\hl{"private\_key\_id": "d4d****4a4"}@*),
"private_key": "-----BEGIN PRIVATE KEY-----\***\n-----END PRIVATE KEY-----\n",
"universe_domain": "googleapis.com"
    \end{lstlisting}
    \caption{Google Cloud service account private key exposed in a certificate file.}
    \label{fig:credential_exposure_google_service_account}
\end{subfigure}

\caption{Examples of credential exposure across different code and configuration formats. As these are sensitive credentials and keys we have obfuscated those.}
\label{fig:credential_exposure_examples}

\end{figure}

\begin{Summary}{Summary of RQ-1}{
\begin{enumerate}
    \item MCP servers exhibit distinct vulnerability patterns compared to other domains of software engineering. Out of eight vulnerability patterns detected in MCP servers, three are common with other domains.
    \item MCP-specific vulnerabilities can be highly prevalent as even an early stage tool could already detect 5.5\% MCP-specific issues.
    \item Credential exposure, e.g., API Keys from FM service providers, GitHub tokens, can cause significant financial loss and major data breaches.
\end{enumerate}
}
\end{Summary}

\subsection*{RQ-2: \rqtwo}\label{sec:rq2}
\motivation
Maintainability remains a pressing concern in modern software systems, especially in the era of FM-based AI applications. According to the \textit{State of Software-2025} report~\cite{state_of_software_2025}, 73\% of AI and big data systems fall below the industry benchmark for maintainability. In mature organizations, overheads to tackle maintainability issues can reach up to \textsterling{250{,}000} (USD~\$332{,}500) per system annually. Two primary sources of those maintainability issues are code smells~\cite{yamashita2012code} and software bugs~\cite{chen2018defects}. Prior research has demonstrated that code smells (e.g., high cognitive complexity) are strongly correlated with increased debugging time and elevated software error rates~\cite{munoz2020empirical}. In addition, many code smells are often linked to bugs or fault-proneness of the systems~\cite{palomba2018diffuseness}, while software bugs themselves impose a considerable engineering burden---up to 39\% of developer effort in leading GitHub projects is spent solely on bug fixing~\cite{amit2021corrective}.

These concerns are particularly relevant to MCP servers, as this is a relatively new software paradigm used to connect FMs with external tools and data sources in comparison to other similar domains, e.g., PyPI or NPM packages. As prior studies suggest that most code smells are introduced during initial code creation rather than being accumulated over time~\cite{tufano2015and}, MCP server implementations may be particularly susceptible to maintainability issues, e.g., code smells, making early assessment critical. Motivated by these risks, this research question aims to systematically characterize the types and prevalence of code smells and software bugs in MCP servers. We also explore whether the prevalence of code smells and bugs has any statistical difference across various programming languages and integration types. By analyzing these systems at an early stage, we seek to provide actionable insights into their maintainability profile and identify patterns that may guide future development and tooling efforts.

\approach
We use the static analysis tool SonarQube to identify both code smells and bugs. We filter and retain only the critical and blocker-level code smells and bugs, and then use an LLM-Jury to categorize these issues into higher-level patterns as described in Section~\ref{sec:exp_design}. We analyze these clusters to understand the most recurrent maintainability issues in the early-stage MCP servers and compare those with other software domains.

In addition, we conduct Kruskal-Wallis H-tests followed by pairwise Mann-Whitney U tests with Bonferroni correction to compare code smell and distributions across different programming languages (e.g., Python, JavaScript, TypeScript) and integration types (official, community, mined). We also compute Cliff’s Delta ($d$) for each pairwise comparison to quantify the effect size.

\findings
We report our findings for code smells and bugs in the following subsections. 

\subsubsection{Prevalence of code smells in MCP server repositories}
\textbf{All high-frequency code smells from traditional and ML systems also appear in MCP servers, indicating that existing techniques can be reused to improve MCP maintainability.} Since different analyzers define code smells at varying levels of granularity (e.g., SonarQube's \textit{high cognitive complexity} vs. PyLint's \textit{line-too-long, complex code}), we align SonarQube-detected smells in MCP servers with top-ranked smells in prior studies on Python ML project~\cite{van2021prevalence} and FM-generated code~\cite{siddiq2022empirical} based on semantic similarity and summarize this in~\tabref{tab:code_smell_types}. To contextualize our findings, we use these semantically similar categories for comparison. We find strong parallels when comparing code smells commonly studied in the JavaScript ecosystem~\cite{saboury2017empirical}. Specifically, top JavaScript smells such as lengthy lines, chained methods, and long functions correspond closely to the \textit{high cognitive complexity} patterns observed in MCP implementations. These overlaps suggest that refactoring techniques from existing literature, e.g., extraction, move~\cite{roy2014vision, singh2018identifying}, de-compositional object techniques~\cite{al2017empirical, singh2018systematic} can also benefit the MCP domain to refactor smelly code.

\begin{table}[ht]
\centering
\small
\caption{Code smell patterns across MCP Servers, ML Projects and FM Generated Code sorted by prevalence in descending order. The first column (MCP server) includes prevalence percentages. Highlighted patterns indicate cross-domain similarities, with superscript
numbers and color denoting the closest semantic match.}
\label{tab:code_smell_types}
\begin{tabularx}{\textwidth}{
  >{\raggedright\arraybackslash}p{5.5cm} 
  >{\raggedright\arraybackslash}p{3.75cm} 
  >{\raggedright\arraybackslash}p{3.55cm}
}
\toprule
\textbf{MCP Servers} & \textbf{ML Projects~\cite{van2021prevalence}} & \textbf{FM-Generated Code~\cite{siddiq2022empirical}} \\
\midrule
\cellcolor{red!20}High Cognitive Complexity (59.7\%)\textsuperscript{\textbf{1}} & \cellcolor{gray!20}unused-wildcard-import\textsuperscript{\textbf{5}} & \cellcolor{orange!50!yellow}Undefined-variable\textsuperscript{\textbf{4}} \\
\cellcolor{orange!20}Code Duplication-Redundancy\textsuperscript{\textbf{2}} (21.4\%) & \cellcolor{red!20}bad-indentation\textsuperscript{\textbf{1}} & \cellcolor{red!20}Line-too-long\textsuperscript{\textbf{1}} \\
\cellcolor{yellow!70}Function Structure Issues (19.4\%)\textsuperscript{\textbf{3}} & \cellcolor{orange!50!yellow}invalid-name\textsuperscript{\textbf{4}} & \cellcolor{yellow!70}Unused-argument\textsuperscript{\textbf{3}} \\
\cellcolor{orange!50!yellow}Variable Declaration and Usage Issues (11.8\%)\textsuperscript{\textbf{4}} & \cellcolor{red!20}line-too-long\textsuperscript{\textbf{1}} & \cellcolor{red!20}Pointless-statement\textsuperscript{\textbf{1}} \\
Asynchronous \& Concurrency Issues (10.8\%) & \cellcolor{yellow!70}missing-function-docstring\textsuperscript{\textbf{3}} & \cellcolor{orange!50!yellow}Pointless-string-statement\textsuperscript{\textbf{4}} \\
Runtime Issues (8.7\%) & \cellcolor{yellow!70}no-member\textsuperscript{\textbf{3}} & \cellcolor{yellow!70}No-member\textsuperscript{\textbf{3}} \\
JavaScript/TypeScript Specific Issues (4.1\%) & \cellcolor{orange!20}duplicate-code\textsuperscript{\textbf{2}} & \cellcolor{orange!50!yellow}Used-before-assignment\textsuperscript{\textbf{4}} \\
Type Safety and Correctness (2.7\%) & \cellcolor{red!20}trailing-whitespace\textsuperscript{\textbf{1}} & \cellcolor{red!20}Superfluous-parenthesis\textsuperscript{\textbf{1}} \\
\cellcolor{gray!20}Import \& Dependency Issues (1.4\%)\textsuperscript{\textbf{5}} &\cellcolor{orange!50!yellow}redefined-outer-name\textsuperscript{\textbf{4}} & \cellcolor{orange!20}Duplicate-code\textsuperscript{\textbf{2}} \\
Python Specific Issues (1.2\%) & \cellcolor{red!20}missing-module-docstring\textsuperscript{\textbf{1}} & \cellcolor{orange!50!yellow}Consider-using-enumerate\textsuperscript{\textbf{4}} \\
\bottomrule
\end{tabularx}
\end{table}

\textbf{66\% of MCP servers contain at least one critical or blocker-level code smell, with some of those code smells, e.g., import \& dependency issues, variable declaration, and usage issues, present in 100\% Python ML projects.}  We identify a total of 17,832 critical and blocker-level code smells across 385 MCP servers and present the distribution of these issues across different programming languages in~\tabref{tab:code_smell_summary_language_wise}. Notably, the median number of critical code smells in MCP servers ranges between 2 and 4 in the most commonly used programming languages, e.g., Python, JavaScript, and TypeScript, while the median number of blocker-level code smells is zero across all these languages. In contrast, traditional software engineering studies have reported that certain code smells can be present in nearly 100\% of the studied Python ML projects and traditional Python projects~\cite{van2021prevalence, chen2016detecting} and up-to 97\% FM generated code can contain code smells~\cite{siddiq2022empirical}.

\textbf{59.7\% of MCP servers suffer from high cognitive complexity, which is also considered as one of the most severe code smells in the Python ecosystem.} As summarized in \tabref{tab:code_smell_types}, we observe that high cognitive complexity is almost three times more prevalent than the second most common code smell, e.g., code duplication-redundancy, in MCP servers. Cognitive complexity is a widely used metric for modeling and estimating the functional complexity, size, and effort required for software development~\cite{wang2013empirical}. While prior studies suggest a threshold of 15 for cognitive complexity~\cite{munoz2020empirical}, violation of this threshold is considered as one of the most severe code smells in the Python ecosystem~\cite{gupta2023severity}. Similarly, this threshold is violated in a substantial portion (59.7\%) of MCP servers, which can lead to increased comprehension time, reduced understandability, and higher debugging time and error rates~\cite{munoz2020empirical}. Prior empirical studies have reported that cognitive complexity positively correlates with comprehension time and subjective ratings of understandability~\cite{munoz2020empirical}. Practitioners need more time to inspect software code when it has low understandability and this has been considered as one of the primary causes of software development cost~\cite{lavazza2023empirical}. Moreover, it can increase the defect proneness of the software system~\cite{antinyan2017evaluating}.

\textbf{In addition to high cognitive complexity, MCP servers also exhibit other recurring code smells that collectively indicate potential long-term maintainability challenges}. The most prevalent of these are code duplication, function structure issues, and variable declaration and usage issues. For instance, in Figure~\ref{fig:code_smell_clone}, we observe repeated timestamp formatting logic across multiple functions within a real-world project, which can cause change misalignment when duplicated segments evolve independently and may lead to the reappearance of the same bug in several places, as also noted in prior software engineering studies~\cite{thummalapenta2010empirical}. Similarly, we observed prevalent function structure issues, such as deeply nested functions, empty methods, or functions that return invariants. These issues collectively increase cognitive burden and obscure program flow, which hinders effective debugging and testing. Prior research attributes excessive nesting to reduced readability and maintainability~\cite{saboury2017empirical}, while invariant-returning or placeholder functions often signal design weaknesses that propagate through the codebase. Variable declaration and usage issues also contribute significantly to maintainability risks~\cite{zabardast2022further}. These include the absence of explicit variable declarations (e.g., missing let or const in JavaScript), exporting mutable module-level variables that can be altered by external modules, and reassignment of function parameters with default values, which can trigger unexpected side effects during repeated invocations. Such practices degrade modularity and complicate reasoning about program state, making MCP servers more prone to subtle and persistent maintenance problems over time.

\begin{figure}[th]
\centering

\begin{subfigure}[t]{0.48\textwidth}
    \vspace{0pt} 
    \begin{lstlisting}[style=mypython, escapeinside={(*@}{@*)}]
def calculate_duration_seconds(start_time: str, end_time: str) -> float:
    """Calculate duration in seconds between two timestamps.

    Args:
        start_time: ISO format timestamp string.
        end_time: ISO format timestamp string.

    Returns:
        float: Duration in seconds.
    """

    if not start_time or not end_time:
        return 0.0

    start = datetime.fromisoformat(
        (*@\hl{\texttt{start\_time.replace("Z", "+00:00") if "Z" in start\_time else start\_time}}@*)
    )
    end = datetime.fromisoformat(
        (*@\hl{\texttt{end\_time.replace("Z", "+00:00") if "Z" in end\_time else end\_time}}@*)
    )
    duration = end - start

    return duration.total_seconds()
    \end{lstlisting}
    \caption{Duration calculation function}
\end{subfigure}
\hfill
\begin{subfigure}[t]{0.48\textwidth}
    \vspace{0pt} 
    \begin{lstlisting}[style=mypython, escapeinside={(*@}{@*)}]
def format_purchase_history(purchases: BillingPurchaseHistoryResponse, limit: int = 5) -> str:
    """Format purchase history into a readable string.

    Args:
        purchases: Billing purchase history response.
        limit: Maximum number of purchase records to include in the output.

    Returns:
        str: Formatted purchase history string.
    """

    if not purchases.purchase_history:
        return "No previous purchases found"

    output = [f"Purchase History (showing {limit} most recent):"]

    for purchase in purchases.purchase_history[:limit]:
        amount = float(purchase.amount) / 100
        (*@\hl{\texttt{timestamp = datetime.fromisoformat(purchase.timestamp.replace("Z", "+00:00"))}}@*)
        formatted_date = timestamp.strftime("%B %d, %Y")
        output.append(f"- ${amount:.2f} on {formatted_date}")

    return "\n".join(output)
    \end{lstlisting}
    \caption{Purchase history formatting function.}
\end{subfigure}

\caption{Example of duplicated timestamp formatting logic appearing across different functions in real-world MCP servers.}
\label{fig:code_smell_clone}

\end{figure}

\begin{table}[t]
\centering
\small
\caption{Distribution of critical and blocker code smells across programming languages. 
\textbf{Critical smell \%} and \textbf{Blocker smell \%} indicate the percentage of projects where at least one critical or blocker-level code smell is present. 
\textbf{Median Critical} and \textbf{Median Blocker} represent the median number of critical and blocker code smells per project, respectively.}
\label{tab:code_smell_summary_language_wise}
\begin{tabular}{lrrrr}
\toprule
\textbf{Language} & 
\textbf{Critical smell \%} & 
\textbf{Median Critical} & 
\textbf{Blocker smell \%} & 
\textbf{Median Blocker} \\
\midrule
Python       & 68.1 & 1.0  & 5.6 & 0 \\
JavaScript   & 39.8 & 2.0    & 1.3 & 0 \\
TypeScript   & 61.1 & 4.0    & 5.8 & 0 \\
Others       & 47.3 & 12.0   & 4.4 & 0 \\
\bottomrule
\end{tabular}
\end{table}

\begin{figure}[tbhp]
    \centering
    \begin{subfigure}[b]{0.46\textwidth}
        \centering
        \includegraphics[width = \textwidth]{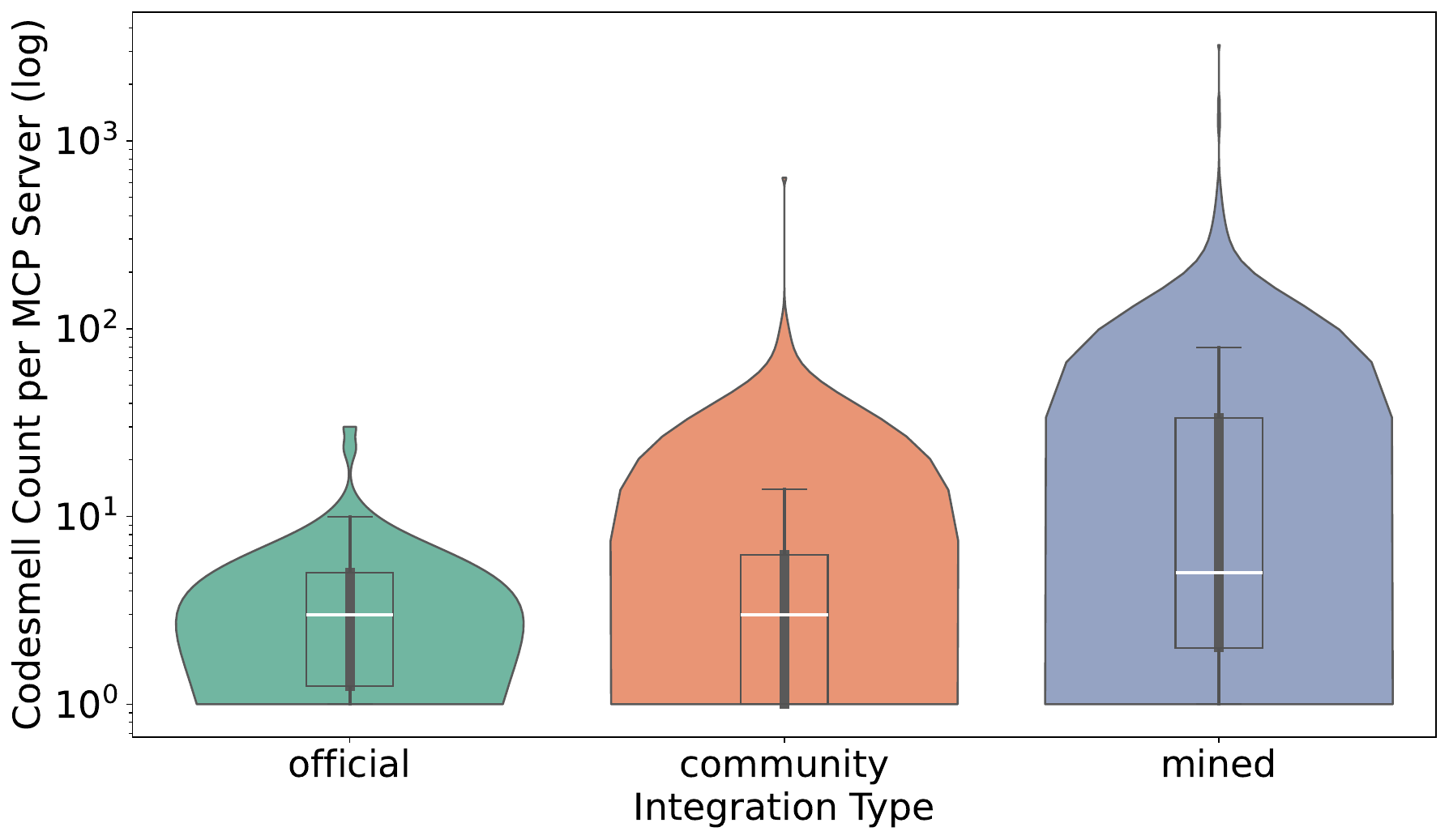}
        \caption{Code smell count distribution by integration type in logarithmic scale.}
        \label{fig:code_smell_violin_integration_type}
    \end{subfigure}
    \hfill
    \begin{subfigure}[b]{0.46\textwidth}
        \centering
        \includegraphics[width=\textwidth]{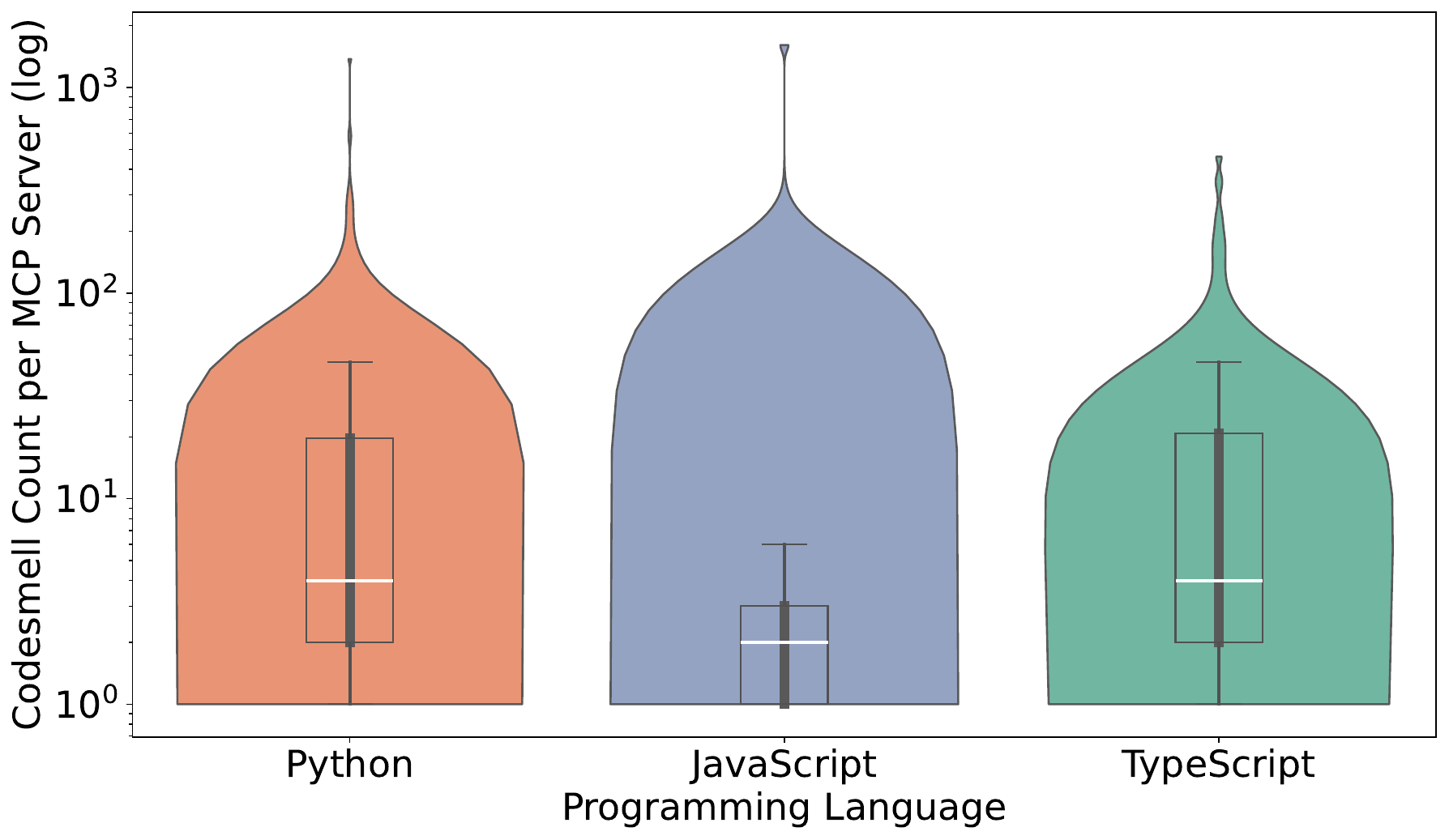}
        \caption{Code smell count distribution by programming language in logarithmic scale.}
        \label{fig:codesmell_language_violin}
    \end{subfigure}
    \caption{Comparison of code smells by integration type and programming language.}
    \label{fig:code_smell_language_violin_combined}
\end{figure}

\textbf{Mined MCP servers contain 66\% more code smells than both official and community servers.} Figure~\ref{fig:code_smell_violin_integration_type} presents the distribution of code smell counts per MCP server across different integration types on a logarithmic scale. We observe a median code smell count of 5 in mined MCP servers, compared to 3 in official and community MCP servers. A Kruskal-Wallis H-test reveals a significant difference in code smell counts among the three integration types ($H = 23.2936$, $p < 0.001$).
Post-hoc Mann-Whitney U tests indicate that mined servers have more code smells than both official ($P-value$ = $0.004$ and $d$ = $-0.322$, \textit{small effect}) and community ($P-value$ = $0.000$ and $d$ = $-0.280$, \textit{small effect}) servers. However, the difference between official and community MCP servers is not statistically significant. We can partially explain by the larger size of mined MCP servers found in Section~\ref{sec:rq0}, as prior work~\cite{yamashita2013code} has shown a positive correlation between code size and code smells.

\textbf{JavaScript MCP servers tend to have 50\% fewer code smells compared to both Python and TypeScript MCP servers.} Figure~\ref{fig:codesmell_language_violin} shows the distribution of code smell counts per MCP server across different programming languages. We observe median code smell counts of 2 in JavaScript MCP servers, compared to 4 in both official and community MCP servers. A Kruskal-Wallis H-test reveals significant differences in code smell counts among the three groups($H = 26.531$, $p < 0.001$). Post-hoc Mann-Whitney U tests show that both Python($Pcorr$ = $0.000$ and $d$ = $0.395$, \textit{medium effect}) and TypeScript($Pcorr$ = $0.000$ and $d$ = $0.415$, \textit{medium effect}) servers have significantly more code smells than JavaScript servers whereas the difference between Python and TypeScript is not statistically significant.

\subsubsection{Prevalence of bugs in MCP server repositories.}
\textbf{Despite language and domain differences, at least 30\% of MCP servers' most frequent bugs overlap with known bug patterns in Java applications, indicating that established debugging techniques can be reused with adaptation.} We report and compare the most frequent bugs in MCP servers with those reported in Java applications analyzed using FindBugs~\cite{ayewah2007evaluating}, as shown in~\tabref{tab:top_bug_types}. While our sample contains no Java-based MCP servers, we used this baseline due to the scarcity of large-scale static bug analyses in Python and JavaScript ecosystems. Unlike vulnerabilities (standardized via CWEs) and code smells (driven by consistent static rules), bugs are influenced by diversified factors, e.g., language and development context~\cite{ray2014large, sztwiertnia2021impact}. Nonetheless, we find three overlapping bug patterns, suggesting partial applicability of conventional bug mitigation strategies in the MCP ecosystem.


\begin{table}[t]
\centering
\small
\caption{Top bug types and their distribution across MCP servers and Java projects. MCP server percentages are shown in parentheses. The table is sorted in descending order of prevalence. Highlighted patterns indicate
cross-domain similarities, with superscript numbers and color denoting the closest semantic match.}
\label{tab:top_bug_types}
\begin{tabularx}{\textwidth}{lX}
\toprule
\textbf{MCP Servers} & \textbf{Java Projects~\cite{ayewah2007evaluating}} \\
\midrule
Array Manipulation Issues (6.2\%) & Nullcheck of value previously dereferenced \\
Parameter Mismatch (3.1\%) & Possible null pointer dereference \\
Type/Structure Issues (2.9\%) & Unwritten field \\
\cellcolor{red!20}Mishandled Exceptions (2.2\%)\textsuperscript{\textbf{1}} & \cellcolor{yellow!50}Invocation of toString on an array\textsuperscript{\textbf{3}} \\
\cellcolor{orange!50!yellow}Object Property and Method Design Issues (1.7\%)\textsuperscript{\textbf{2}} & Class defines field that masks a superclass field \\
\cellcolor{yellow!50}Iterable and Collection Issues (1.7\%)\textsuperscript{\textbf{3}} & \cellcolor{orange!50!yellow}Method call passes null for unconditionally dereferenced parameter\textsuperscript{\textbf{2}} \\
Infinite Loop (1.7\%) & \cellcolor{red!20}Possible null pointer dereference in method on exception path\textsuperscript{\textbf{1}} \\
Style Issues (0.9\%) & Method ignores return value \\
Getter/Setter and Object Property Access (0.7\%) & Field only ever set to null \\
 & Suspicious reference comparison \\
\bottomrule
\end{tabularx}
\end{table}

\textbf{We identify nine distinct bug categories affecting 14.4\% of MCP servers, all of which have been studied before in Java, Python or JavaScript ecosystems}. From~\tabref{tab:top_bug_types}, we observe that the top three bug patterns in MCP servers, i.e., \texttt{Array Manipulation Issues}, \texttt{Parameter Mismatches}, and \texttt{Type/Structure Issues}, do not appear in the Java ecosystem's reported bug taxonomies. Of the other six bug categories, three (i.e., \texttt{Mishandled Exceptions}, \texttt{Collection Misuse}, and \texttt{Object Method Design Issues}) have been spotted before in the Java ecosystem. Those not found in the Java ecosystem (including the most prevalent three), 
do occur in other ecosystems, and hence are known to practitioners and the research community. For example, \texttt{Array Manipulation Issues}, which often arise when practitioners do not pass any comparator function while sorting an array, can lead to unintended behavior such as lexicographic sorting in the JavaScript community~\footnote{\url{https://v8.dev/blog/array-sort}}. Other than that, parameter mismatch where a function is called with incorrect arguments in terms of name or number~\cite{rice2017detecting}, type/structure issues originating from wrong type intersection~\cite{castagna2017gradual}, infinite loop~\cite{wang2019loopfix}, style issues related to invalid CSS property~\cite{park2015analysis} and getter/setter and object property access issue caused by unintentional write or read behavior in getter/setter methods~\cite{arteca2023statistical} have been identified in the Python and JavaScript literature. These overlaps suggest that existing debugging practices can address many MCP-specific bugs, at least at the detection and fix level.

\begin{figure}[t]
    \
        \begin{subfigure}[b]{0.48\textwidth}
        \centering
        \includegraphics[width=\textwidth]{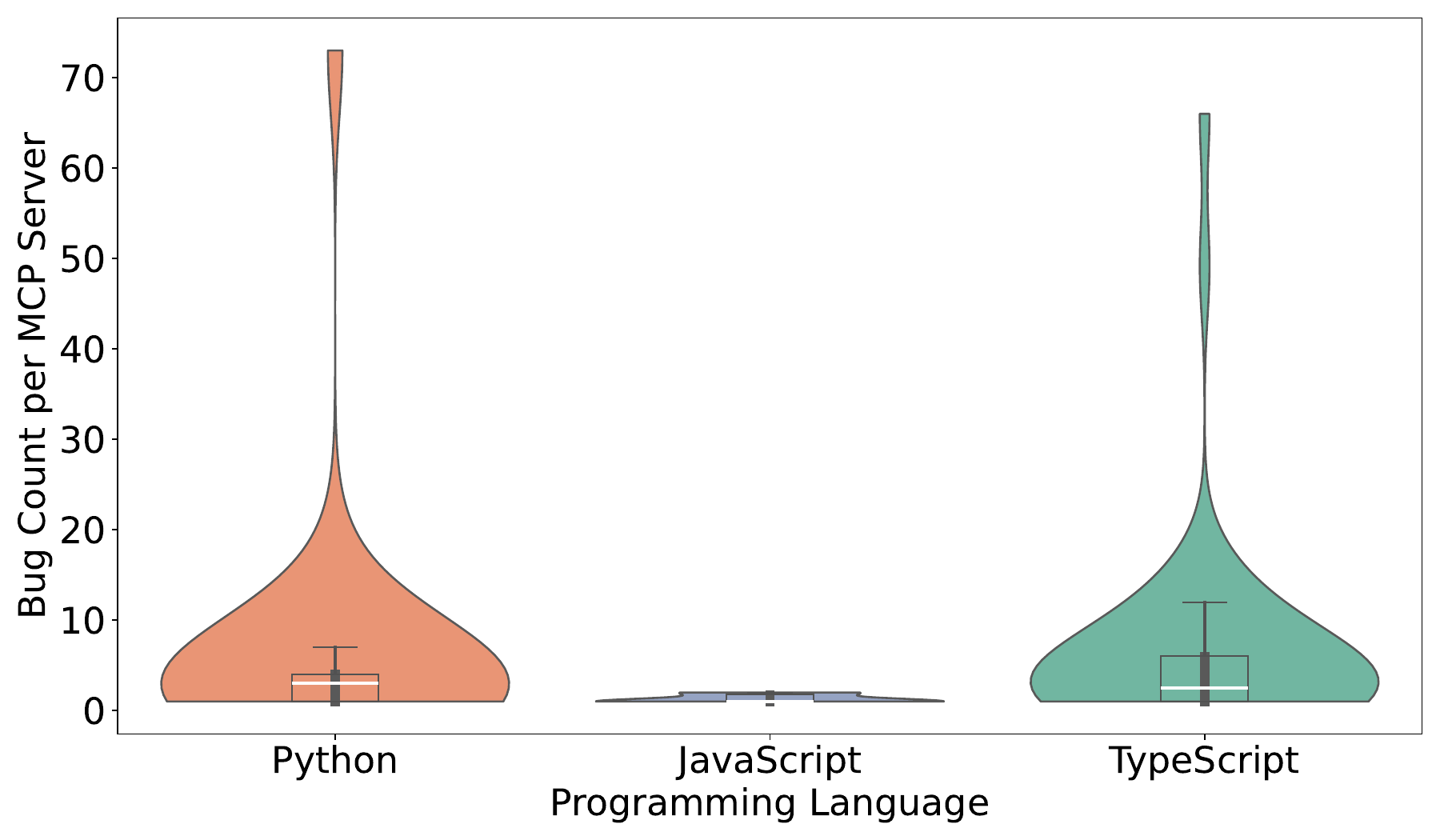}
        \caption{Bug distribution per MCP server grouped by language.}
        \label{fig:bug_language_violin}
    \end{subfigure}
    \hfill
    \begin{subfigure}[b]{0.48\textwidth}
        \centering
        \includegraphics[width = \textwidth]{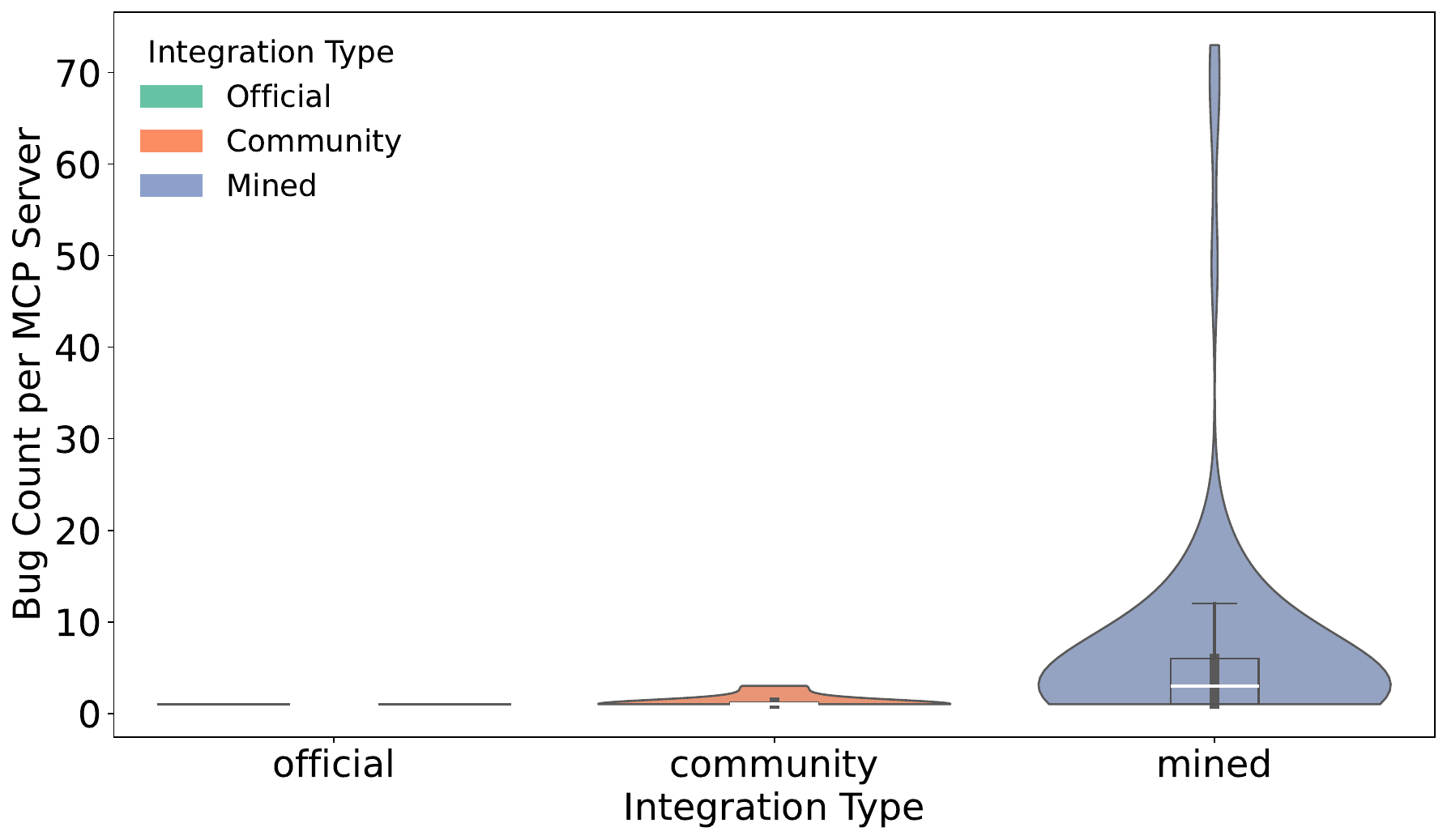}
        \caption{Bug  distribution per MCP server grouped by Integration Type.}
        \label{fig:bug_violin_integration_type}
    \end{subfigure}
    \caption{Comparison of bugs by integration type and programming language across MCP servers.}
    \label{fig:bug_language_violin_combined}
\end{figure}

\begin{table}[t]
\centering
\small
\caption{Distribution of critical and blocker bugs across programming languages. 
\textbf{Critical bug \%} and \textbf{Blocker bug \%} indicate the percentage of projects where at least one critical or blocker-level bug is present. 
\textbf{Median Critical} and \textbf{Median Blocker} represent the median number of critical and blocker bugs per project, respectively.}
\label{tab:bug_summary_language_wise}
\begin{tabular}{lrrrr}
\toprule
\textbf{Language} & 
\textbf{Critical bug \%} & 
\textbf{Median Critical} & 
\textbf{Blocker bug \%} & 
\textbf{Median Blocker} \\
\midrule
Python       & 3.6  & 0   & 8.1 & 2.0 \\
JavaScript   & 2.6  & 2.0   & 1.5 & 0 \\
TypeScript   & 13.1 & 3.0   & 1.8 & 0 \\
Others       & 7.0  & 46.0  & 3.0 & 3.0 \\
\bottomrule
\end{tabular}
\end{table}
\emp{Bug percentages in MCP servers do not vary across different programming languages whereas mined MCP servers contain 67\% more bugs than community MCP servers}. We detect a total of 523 bugs across 84 unique MCP projects. We illustrate the language-wise bug distribution in \tabref{tab:bug_summary_language_wise} and also the corresponding violin plot distributions in Figure~\ref{fig:bug_language_violin}. Although from the table it appears that TypeScript can be more susceptible to critical bugs and Python is more susceptible to blocker bugs, a Kruskal-Wallis statistical test indicates that there is no significant difference in the prevalence of bugs with respect to the programming language of MCP servers ($H = 4.9$, $p = 0.178$). 

In contrast, bug prevalence across integration types as shown in Figure~\ref{fig:bug_violin_integration_type}. A Kruskal-Wallis H-test reveals significant differences in bug counts among the three groups with ($H = 9.6$, $p = 0.008$). Post-hoc Mann-Whitney U tests with Cliff's delta show that mined servers have significantly more bugs than community servers ($p-value$ = $0.026$ and $d$ = -$0.557$, \textit{large effect}) where the differences in other pairs are not significant. We suspect that the higher commit activity reported in Section~\ref{sec:rq0} can have an association with this behavior as previous studies indicated that a higher frequency of commits, especially those not adhering to contribution norms, was associated with an increased likelihood of introducing bugs~\cite{falcao2020relating}.


\begin{Summary}{Summary of RQ-2}{
\begin{enumerate}
    \item 66\% of MCP servers contain at least one critical or blocker-level code smell, and 14.4\% exhibit at least one statically detected bug, indicating widespread maintainability concerns.
    \item The most prevalent code smell is high cognitive complexity, affecting 59.7\% of MCP servers---three times more frequent than the next most common smell---and strongly linked to reduced understandability and higher error risk.
    \item Mined MCP servers exhibit more code smells and bugs than official or community servers, possibly due to their higher development activity or less structured maintenance practices.
    \item All identified MCP bug types overlap with known bugs from Java, Python, and JavaScript ecosystems, suggesting that existing debugging and refactoring tools may be reused or adapted for MCP development.
\end{enumerate}
}
\end{Summary}

\section{Implications}\label{sec:implications}
MCP servers demonstrated a strong indication of long-term sustainability regarding different metrics we studied in Section~\ref{sec:rq0}. Hence, we must make sure the MCP ecosystem grows in a secure and maintainable manner. In this section, we discuss the implications of our study on the security and maintainability dimensions of MCP servers. Specifically, we have three target audiences in the MCP ecosystem: (i) Researchers, (ii) Practitioners, and (iii) Ecosystem, e.g., MCP Registry, Maintainers. In the following subsections, we discuss the implications of our findings for each of these target audiences.

\subsection{Implication for Researchers}

\textbf{Security researchers should expand standard vulnerability databases, e.g., CWE, OWASP, to include MCP-specific threats.} Our study reveals that common MCP vulnerabilities, e.g., such as tool poisoning, credential exposure, and transport security issues, are largely absent from existing vulnerability taxonomies (e.g., CWE, OWASP). While we observed 13 distinct CWEs in MCP servers, only one (CWE-306) appears in the current MITRE Top 25. On the other hand, while we have OWASP top-10 for GenAI\footnote{\url{https://genai.owasp.org/llm-top-10/}}, we do not have anything similar for the MCP domain. We urge the security research community to formalize MCP-specific threat models and advocate for their inclusion in standard databases and taxonomies.

\textbf{Researchers should develop MCP-specific vulnerability analysis tools to uncover hidden risks beyond what general-purpose scanners can detect.} While traditional tools like SonarQube successfully identify some MCP vulnerabilities, e.g., credential exposure and transport security issues, our findings showed that these issues appear in only a small fraction of repositories. This low prevalence suggests an unusually secure ecosystem or, more likely, an under-detection problem rooted in the limitations of current tools. Additionally, existing MCP-specific tools like mcp-scan remain in early development stages and currently target only one class of vulnerability (tool poisoning), with limited reliability and coverage. These gaps underscore the need for new MCP-specific tools that account for the unique operational characteristics of MCP servers---including dynamic tool invocation, runtime dependency management, and FM-specific interaction patterns. Inspired by our hybrid analysis approach, future research can explore integrated frameworks that combine static and dynamic analysis, program instrumentation, and FM-aware code inspection to surface latent vulnerabilities specific to the MCP domain.

\subsection{Implication for Practitioners}
We can classify MCP practitioners into two groups : (i) MCP Developers who build and manage MCP servers and tools and (ii) MCP Users who use MCP servers to build FM-based AI applications. Each group can benefit by interpreting the results of this study and applying those in real-life scenarios.

\textbf{MCP developers must adopt proactive security practices to prevent credential exposure and related vulnerabilities.} Our findings reveal that sensitive credentials, e.g., API keys, service account secrets, and access tokens, are frequently left exposed in MCP server codebases (Section~\ref{sec:rq1}). These exposures pose immediate threats, including unauthorized system access, service abuse, and financial losses. Developers should integrate static application security testing (SAST) into their CI pipelines and adopt secure coding practices to mitigate these risks, as suggested in prior research~\cite{cruz2023open}. While tools like SonarQube can detect many low-hanging issues, we recommend extending this with MCP-specific tools (e.g., mcp-scan) to catch issues unique to the domain. In addition, teams should implement secure development workflows incorporating peer-reviewed pull requests, automated secret detection (e.g., GitLeaks), and environment-based credential management to reduce exposure risks~\cite{prates2025devsecops,basak2022practices}.

\textbf{MCP developers should leverage proven software engineering practices alongside emerging ML and LLM-based techniques to detect and remediate code smells and bugs.} Our study finds that MCP servers exhibit code smells and bugs that closely mirror those observed in traditional software domains (Section~\ref{sec:rq2}). This similarity allows practitioners to apply well-established techniques from the broader software engineering literature. Prior studies demonstrate that adhering to established design patterns can reduce the proneness to code smells~\cite{alfadel2020empirical}, while specific refactoring techniques (e.g., \textit{Extract Method}, \textit{Pull Up Method}) can help to eliminate existing smells~\cite{lacerda2020code,fowler2018refactoring}.  Complementing these traditional methods, recent works demonstrate that FMs can identify smells, generate human-like refactorings, and even provide semantic bug fixes~\cite{alomar2024refactor, wu2024ismell}. 

For MCP projects in particular, this opens several practical directions: developers may embed AI-assisted analyzers into their pre-commit hooks to automatically surface code smells and static bugs, rely on intelligent coding assistants within development environments (IDEs) to flag problematic constructs as code is written, and leverage automated refactoring tools to improve legacy components and reduce long-term technical debt systematically. Given the structural and behavioral parallels between MCP servers and traditional software systems, developers are encouraged to integrate these advanced tools into their workflows. From a scalability perspective, these interventions are relatively low overhead because they operate directly on source code and can be integrated into existing IDEs without requiring live server execution.

\textbf{MCP users should audit third-party MCP servers for security vulnerabilities before deployment to avoid downstream risk propagation.} Our findings indicate that vulnerabilities from MCP servers, e.g., lack of access control, improper resource management, transport security issues, and insecure file creation, may propagate into downstream applications and compromise the host environment. In addition, emerging threats like tool poisoning introduce new risks whereby FMs could be coerced into executing malicious toolchains on user infrastructure, leading to persistent compromise or data exfiltration. To mitigate these risks, MCP users should incorporate proactive auditing workflows. While still evolving, tools such as SonarQube and mcp-scan offer a foundational level of detection and can surface common vulnerabilities. Although these tools may not detect the full spectrum of MCP-specific threats, our study shows that they can still identify some key issues, which makes them a good starting point. MCP users should have this bare minimum audit process until the research community develops better, state-of-the-art tools. 

In operational settings, users must also account for deployment complexity and runtime overhead. mcp-scan requires a live MCP server instance, a valid client configuration, and often real service credentials. Repeated dynamic scans can become resource-intensive when servers depend on external APIs, interactive setup steps, or rate-limited services. To address these scalability challenges, organizations can adopt a staged auditing strategy: (1) perform lightweight static scans and secret detection across all candidate servers;(2) execute dynamic mcp-scan analysis only for shortlisted or high-risk servers; and(3) conduct dynamic rescans on a scheduled basis (e.g., during low-traffic windows) or when tool descriptions, permissions, or dependencies change. Additionally, sandboxed or containerized testing environments can be used to isolate dynamic scans, thereby reducing performance impact on production systems and limiting the exposure of real credentials during evaluation.

This hybrid scanning approach combines traditional static analysis with MCP-specific dynamic analysis, which is generalizable across sectors because MCP servers expose two distinct attack surfaces: (1) conventional software vulnerabilities in executable code and (2) protocol-specific threats embedded in reflective metadata.

In financial services, practitioners deploying payment processing MCP servers (such as the Stripe integration presented in our motivating example) can apply static analysis to inspect source code for hardcoded credentials, insecure API key storage, and improper error handling in transaction logic. In parallel, MCP-scan can examine runtime tool descriptions to detect poisoning attempts that could manipulate payment recipients or transaction amounts. 

In healthcare, where MCP servers connect to electronic health record (EHR) systems and diagnostic databases, the same hybrid approach applies. Static scanners can audit code for access control flaws, insecure database queries, and improper resource cleanup that could leak protected health information (PHI), while dynamic scanning validates that tool descriptions do not contain instructions designed to coerce foundation models into unauthorized patient data access or exfiltration. 

In software development, particularly when integrating infrastructure MCP servers like Microsoft's Azure connector mentioned in our introduction, static analysis detects insecure file creation, command injection vulnerabilities, and transport security issues in deployment automation code. At the same time, mcp-scan inspects tool metadata to prevent poisoned descriptions from redirecting repository operations, compromising CI/CD pipelines, or leaking source code through manipulated tool parameters. Thus, by integrating source-level inspection with runtime reflection analysis, our approach provides a domain-agnostic yet sector-aware security validation pipeline that scales across real-world MCP deployments.

\subsection{Implication for Ecosystem Maintainers}

\textbf{Ecosystem (e.g., MCP registry) maintainers should establish standardized governance mechanisms and provide clear developer guidelines to ensure baseline security and quality}. MCP registries, e.g., Smithery~\cite{smithery}, Glama~\cite{glamma}, or Cloudflare Workers~\cite{cloudflare_mcp}, serve as centralized hubs for discovering and deploying MCP servers. As our study already highlighted security vulnerabilities and maintainability issues in MCP servers, this calls for broader governance initiatives from registry maintainers. While some registries implement basic review processes (e.g., isolated manual testing or simple pull requests), these are often limited to infrastructure setup or package-level scans. For instance, Glama performs manual review in an isolated VM and scans the server for Python and NPM dependencies\footnote{\url{https://www.reddit.com/r/mcp/comments/1hm3g2s/glama_mcp_server_directory/}} and Smithery performs automated analysis of the MCP server code and raises simple pull requests to create Dockerfiles or any obvious missing dependencies. However, these checks fail to capture issues like credential exposure or MCP-specific tool poisoning. To mitigate these risks, registries should formalize governance protocols, including submission checklists, required permission disclosures, and security guideline conformance similar to other mature registries, e.g., Google Play Store or iOS app store~\cite{ios-guideline, android-guideline}. 

\textbf{MCP registries should incorporate automated vulnerability and sustainability scanning and revocation workflows to prevent the propagation of insecure MCP servers.} As centralized distribution platforms, registries are critical in ecosystem-wide security. However, without automated pre-publication scanning and post-deployment monitoring, they risk serving as vectors for vulnerability propagation. Inspired by mature ecosystems like the Google Play Store, Docker Hub, and the iOS App Store, registry maintainers should establish continuous monitoring and enforce automated revocation policies for non-compliant or vulnerable servers~\cite{calciati2017apps,wang2018android}. These policies and workflows reduce the burden on end users and create a scalable security model that will evolve with the growing adoption of MCP infrastructure.

\section{Threats to Validity}\label{sec:threadstovalidity}





\subsection{External validity}
In this study, we analyzed open-source MCP servers, acknowledging that they may not fully represent the entire population of MCP servers. To mitigate this threat to external validity, we constructed our dataset using a diverse set of sources, including Anthropic's official curated list and additional MCP servers mined from GitHub using a search-based approach. 

Similarly, the vulnerabilities identified through our SonarQube analysis may not constitute an exhaustive list of all potential vulnerabilities in the MCP servers. Nevertheless, to the best of our knowledge, SonarQube’s security rules cover all categories outlined in the MITRE CWE Top 25, OWASP Top 10, and PCI DSS standards~\cite{sonarqube_security_rules}. Therefore, we expect that the most prevalent and critical software vulnerabilities are reflected in our analysis. Additionally, we utilized \texttt{mcp-scan} to detect MCP-specific vulnerabilities (e.g., tool poisoning). While \texttt{mcp-scan} is a relatively new and continuously evolving tool, it was the most widely recognized and actively maintained option available at the time of our study.

We report the prevalence of code smells and bugs based on SonarQube analysis in RQ-2. Although prior research has noted that SonarQube-reported ``bugs'' may not always correspond to actual faults, and that some ``code smells'' may exhibit a stronger correlation with real defects~\cite{lenarduzzi2020sonarqube}, our focus in RQ-2 is on the maintainability of MCP servers. Accordingly, we report both bugs and code smells as indicators of maintainability concerns.

Also, as our mining technique heavily relies on the import patterns of official MCP SDKs, repositories using MCP without explicit imports, e.g., via other third-party SDKs, may be missed from our analysis. However, as the primary mining could gather 1,899 MCP servers, the search patterns of official MCP SDKs are sufficient at the time of the study.

Another potential threat to validity stems from the fact that mcp-scan was applied to a statistically representative subset rather than the full set of 583 MCP servers. Because dynamic analysis with mcp-scan requires complete server installation, configuration with valid credentials, and live runtime interaction, conducting such analysis at scale across all repositories was operationally infeasible. To mitigate this limitation, we determined the sample size using Cochran’s formula with finite population correction under a 95\% confidence level and 10\% margin of error, resulting in 83 randomly selected servers. Under these statistical assumptions, the reported prevalence of MCP-specific vulnerabilities is expected to approximate the true ecosystem distribution within the specified error bounds, thereby providing a realistic and statistically grounded estimate of dynamic risk.

Finally, mcp-scan relies on the reflection protocol (as discussed in~\secref{background-reflection}) to retrieve and analyze tool descriptions through its proprietary invariant guardrailing API. At the time of this study, the tool does not inspect source code or execution behavior. Rather, mcp-scan identifies potential vulnerabilities based on the structural integrity and semantic content of tool descriptions. Consequently, its detection capability may not capture deeper code-level or runtime-specific flaws. For example, it does not perform advanced dynamic testing, e.g., fuzzing, runtime monitoring, or execution-path exploration, which may reveal adaptive or context-dependent risks. To the best of our knowledge, no MCP-specific tools currently exist to support such comprehensive dynamic analysis. However, we complemented the mcp-scan with a static analyzer such as SonarQube, so that our framework jointly covers code-level issues identified by static analysis and dynamic vulnerabilities that can compromise the runtime behavior of the tools, achieving the best overall coverage with the techniques available at the time of this research.

\subsection{Construct validity}
As MCP servers represent a nascent technology space, many open-sourced projects may be experimental or built by early adopters. Though we applied a star-based filter to exclude the toy projects, we acknowledge previous works stating that star count may not be a good proxy for repository quality or relevance~\cite{munaiah2017curating}. However, as the MCP concept itself is less than a year old, we could not go for a more rigorous filtering and chose a 10-star threshold as a reasonable compromise between the data quality and the quantity, as suggested in other previous works~\cite{dabic2021sampling}.

The metrics used in this study to assess future sustainability, as discussed in Section~\ref{sec:rq0}, may not represent an exhaustive set of possible indicators. Additionally, for some metrics, median values were not available in other domains. Nevertheless, we made extensive efforts to gather metric definitions from existing literature~\cite{xiao2023early, coelho2017modern} and reviewed all possible empirical studies to obtain corresponding values. It is important to note that our goal in presenting these metrics across different ecosystems is not to offer an absolute comparison, but rather to provide an indication of whether an ecosystem appears to be in a generally healthy state.

Another threat to construct validity lies in the use of an LLM-Jury for clustering, as the process could be susceptible to errors such as hallucination, bias, or misclassification. We implemented a multi-faceted mitigation strategy to address this threat. First, to ensure our generated patterns are conceptually sound and not hallucinations, we grounded them by aligning them with established Common Weakness Enumerations (CWEs), a standard practice for validating vulnerability analyses~\cite{rahman2019seven, verdi2020empirical}. Second, to validate the classification accuracy and control for potential biases inherited from LLM training data, we performed a manual verification on a random sample of 75 issues across all categories. As detailed in~\secref{validating-issue-patterns-manually}, the near-perfect agreement between the LLM-Jury and human experts provides strong evidence that systematic biases or significant misclassifications did not materially affect our results. Finally, to address the threat of randomness, we evaluated the consistency of the Jury's output across multiple runs (\secref{evaluating-the-consistency-of-llm-jury}). The high consistency of cluster membership confirms that our methodology is robust and that the resulting patterns are stable. Together, these mitigation steps provide high confidence in the validity of the issue patterns derived from our LLM-Jury system.

\subsection{Internal validity}
In this study, we investigated open-source MCP servers' health, sustainability, security vulnerabilities, and maintainability issues. A key methodological challenge lies in establishing appropriate baselines for comparison, especially given the young age of the MCP ecosystem.  

For RQ-0, where we studied development and community health, we derived baselines from a broad set of prior works focused on general open-source projects. These metrics are widely referenced in prior literature~\cite{xiao2023early, coelho2017modern}, but no single prior study provided a unified baseline covering all metrics considered (e.g., commit frequency, CI adoption, contributor count). Hence, we synthesized baselines by aggregating metric distributions reported across diverse studies with varying sampling criteria, project maturity, and domains. For example, metrics for general OSS projects often reflect mature repositories with multi-year histories, while MCP servers are, on average, less than six months old. To mitigate this, we normalize time-dependent measures, perform statistical tests among the integration types and highlight relative positioning rather than making absolute judgments.

In RQ-1, we compared MCP vulnerabilities with those reported in PyPI, NPM, and infrastructure-as-code (IaC) ecosystems. However, the methodologies of these baseline studies differ from ours. For example, PyPI and NPM vulnerability counts are drawn from third-party vulnerability databases (e.g., Snyk), which rely on community reporting. In contrast, we employ static analysis via SonarQube and MCP-specific tooling. The prior IaC study~\cite{rahman2019seven} partially aligns methodologically through static analysis. To mitigate this, we also reported the static analysis focusing vulnerabilities~\cite {ruohonen2021large, latendresse2022not}, which unfortunately reports only prevalence and does not prioritize taxonomy. As the motivation of this study is to understand the current state-of-the-art landscape of MCP servers without a comparable baseline, it is hard to relate the findings that justify the baseline choices.

In RQ-2, the bug taxonomy we use for comparison is drawn from FindBugs-based studies on Java applications~\cite{ayewah2007evaluating}. Though none of the MCP servers analyzed are written in Java, we could not find any comprehensive, static-analysis-driven bug taxonomies for Python, JavaScript, or TypeScript---the dominant languages in the MCP ecosystem. Hence, using Java-based taxonomies as reference points represents an unavoidable compromise due to the limited availability of domain-specific baselines. To mitigate this, we also studied language-specific research on individual bug patterns observed in MCP servers and reported those findings separately to address this.

Finally, MCP is a rapidly evolving paradigm, and our baseline choices reflect the best available options at the time of study. As the ecosystem matures and more empirical research becomes available, we expect the relevance and precision of comparisons to improve. 
\section{Conclusion}\label{sec:conclusion}
Our paper presents the first large-scale empirical study of 1,899 MCP servers, evaluating their sustainability, security, and maintainability using a hybrid analysis approach that combines a general-purpose static analysis tool (i.e., SonarQube) with an MCP-specific tool (i.e., \texttt{mcp-scan}).

Our findings show that MCP servers generally exhibit strong signals of long-term sustainability, with higher or equal median values in 9 out of 14 key development and community metrics. Maintainability issues such as code smells and bugs---affecting 66\% and 14.4\% of MCP servers, respectively---occur at rates comparable to traditional open-source systems. Our findings suggest that established techniques for assessing sustainability and improving maintainability remain applicable in the MCP context.

In contrast, MCP servers diverge notably in their security profile. We identify eight distinct vulnerability patterns in 7.2\% of MCP servers, only three of which overlap with common issues in ecosystems like Python or Infrastructure-as-Code. Additionally, 5.5\% of MCP servers exhibit tool poisoning---an emerging, MCP-specific threat.

These findings signal a shift in the risk surface of modern FM-based AI application infrastructure and highlight the need for dedicated tools and taxonomies to identify and address MCP-specific vulnerabilities. As MCP adoption accelerates, researchers, practitioners, and registry maintainers must invest in domain-specific security tooling, automated auditing, longitudinal tracking of vulnerability patches, and robust governance to ensure the safe and reliable evolution of FM-based software systems.



\section*{Disclaimer}
Any opinions, findings, and conclusions, or recommendations expressed in this material are those
of the author(s) and do not reflect the views of Huawei. We have used AI tools, e.g., Grammarly, ChatGPT, Gemini in this study to fix the grammatical and writing issues.

\bibliographystyle{ACM-Reference-Format}
\bibliography{main.bib}
\appendix

\section{Appendix}\label{sec:appendix}

\subsection{Prompts used by the LLM-Jury}\label{prompt-jury}

\subsubsection{Worker Prompt}

\begin{tcolorbox}[breakable, enhanced, colback=white, colframe=black,
  boxrule=0.4pt, sharp corners, fontupper=\footnotesize]

\textbf{\# TASK}\\
Analyze the SonarQube vulnerabilities provided below and organize them into meaningful security clusters. 
Group vulnerabilities that share similar root causes, require similar remediation approaches, or address 
related security concerns.

\vspace{1em}
\textbf{\# CLUSTERING GUIDELINES}

\textbf{Criteria for Grouping}
\begin{itemize}
    \item \textbf{Root Cause Similarity}: vulnerabilities originating from the same weakness
    \item \textbf{Remediation Approach}: fixed using similar techniques
    \item \textbf{Security Domain}: authentication, encryption, injection, etc.
    \item \textbf{Industry Standards Alignment}: consider OWASP Top 10 and CWE taxonomy
\end{itemize}

\textbf{Cluster Characteristics}
\begin{itemize}
    \item Create clusters at appropriate granularity.
    \item Each cluster should contain at least 2--3 vulnerabilities.
    \item Cluster labels should be:
    \begin{itemize}
        \item Clear and descriptive (2--5 words)
        \item Use professional security terminology
        \item Actionable and specific
        \item Consistent naming style
    \end{itemize}
\end{itemize}

\textbf{Avoid}
\begin{itemize}
    \item Overly broad clusters (e.g., ``Security Issues'')
    \item Overly granular clusters (one vulnerability per cluster)
    \item Mixing unrelated domains
    \item Redundant or overlapping clusters
\end{itemize}

\vspace{1em}
\textbf{\# OUTPUT FORMAT}

\begin{verbatim}
| Violated Rule Name | Quality Impact | Count | Cluster Label |
|--------------------|----------------|-------|---------------|
| [Rule name]        | [SECURITY/...] |   #   | [Cluster]     |
\end{verbatim}

\vspace{1em}
\textbf{Cluster Summary}

After the table, provide a short summary for each cluster:

\begin{verbatim}
[Cluster Label]: 1–2 sentence description
\end{verbatim}

\vspace{1em}
\textbf{\# INPUT DATA}\\
Your vulnerability data here

\end{tcolorbox}

\subsubsection{Judge Prompt}

\begin{tcolorbox}[breakable, enhanced, colback=white, colframe=black,
  boxrule=0.4pt, sharp corners, fontupper=\footnotesize]

\textbf{\# ROLE}\\
You are an expert code quality analyzer and security specialist tasked with evaluating vulnerability clustering results and assigning the most appropriate cluster label to each vulnerability.

\vspace{1em}
\textbf{\# TASK}\\
You will receive:
\begin{enumerate}
    \item Original dataset: SonarQube vulnerabilities with their Quality Impact types and counts.
    \item Model A's clustering: cluster labels assigned by Model A to each vulnerability.
    \item Model B's clustering: cluster labels assigned by Model B to each vulnerability.
\end{enumerate}

Your task is to evaluate both models' cluster assignments for each vulnerability and determine the most appropriate cluster label for each row.

\vspace{1em}
\textbf{\# EVALUATION CRITERIA}

For each vulnerability, assess the cluster labels based on:

\begin{enumerate}
    \item \textbf{Semantic Accuracy (40\%)}  
    Which label best represents the vulnerability's nature?  
    Does the label capture the core security or quality issue?  
    
    \item \textbf{Consistency (30\%)}  
    Does the label group this vulnerability with semantically similar issues?  
    Is the categorization aligned with OWASP, CWE, or widely recognized security taxonomies?
    
    \item \textbf{Clarity (20\%)}  
    Is the label clear, meaningful, and actionable for developers?  
    Does it avoid ambiguity?
    
    \item \textbf{Remediation Alignment (10\%)}  
    Would this grouping help teams address similar root causes together?
\end{enumerate}

\vspace{1em}
\textbf{\# DECISION RULES}
\begin{itemize}
    \item Choose Model A's label if it is clearly superior.
    \item Choose Model B's label if it is clearly superior.
    \item If both are equally valid, prefer the more specific or descriptive label.
    \item If both are inadequate, create a new label that better represents the vulnerability.
    \item Maintain consistency: if you assign a label to one vulnerability, use the same label for similar vulnerabilities.
\end{itemize}

\vspace{1em}
\textbf{\# OUTPUT FORMAT}

\begin{verbatim}
| Vulnerability Name | Quality Impact | Count | Model A Label | Model B Label | Final Label | Reasoning |
|-------------------|----------------|-------|---------------|---------------|-------------|-----------|
| [Name] | [Impact] | [Count] | [Label A] | [Label B] | [Your Choice] | [Brief explanation] |
\end{verbatim}

\vspace{1em}
\textbf{Summary Statistics}
\begin{verbatim}
- Total vulnerabilities: [X]
- Unique final clusters: [Y]
- Agreed with Model A: [X times]
- Agreed with Model B: [X times]
- Created new labels: [X times]
\end{verbatim}

\vspace{1em}
\textbf{Final Cluster Definitions}
\begin{verbatim}
[Cluster Label 1]: [Brief description]
[Cluster Label 2]: [Brief description]
...
\end{verbatim}

\vspace{1em}
\textbf{\# GUIDELINES}
\begin{itemize}
    \item Use clear, professional terminology (e.g., Authentication \& Authorization, Input Validation, Sensitive Data Exposure).
    \item Prioritize semantic meaning over simply choosing one model.
    \item Maintain balanced granularity (typically 5--15 clusters).
    \item In the Reasoning column, give a concise one-sentence justification.
    \item If creating new labels, ensure they are not redundant with existing categories.
\end{itemize}

\vspace{1em}
\textbf{Sample Input Format Template}

\begin{verbatim}
# INPUT DATA

## Original Dataset
| Vulnerability Name | Quality Impact | Count |
|-------------------|----------------|-------|
| PostgreSQL database passwords should not be disclosed | SECURITY | 16 |
| Origins should be verified during cross-origin communications | SECURITY | 11 |
| ... | ... | ... |

## Model A Clustering
| Vulnerability Name | Cluster Label |
|-------------------|---------------|
| PostgreSQL database passwords should not be disclosed | Credential Management |
| Origins should be verified during cross-origin communications | Web Security |
| ... | ... |

## Model B Clustering
| Vulnerability Name | Cluster Label |
|-------------------|---------------|
| PostgreSQL database passwords should not be disclosed | Secrets & Credentials Exposure |
| Origins should be verified during cross-origin communications | Cross-Origin Security Issues |
| ... | ... |
\end{verbatim}

\end{tcolorbox}

\subsection{Evaluating the Consistency of the Issue Patterns Identified by LLM-Jury}\label{jury-consistency-details}

\small
\setlist[itemize]{leftmargin=*, nosep}

\begin{longtable}{|
  p{0.32\textwidth}|
  p{0.50\textwidth}|
  >{\raggedleft\arraybackslash}p{0.10\textwidth}|
}
\caption{Consistency of vulnerability clusters produced by the LLM-Jury across five independent runs.}
\label{tab:cluster_consistency} \\
\hline
\textbf{Cluster Names in Different Runs of Vulnerabilities} &
\textbf{Issues Under the Cluster} &
\textbf{Cluster Naming Consistency} \\
\hline
\endfirsthead

\hline
\textbf{Cluster Names in Different Runs of Vulnerabilities} &
\textbf{Issues Under the Cluster} &
\textbf{Cluster Naming Consistency} \\
\hline
\endhead

\hline
\endfoot

\hline
\endlastfoot

\textbf{Run1}: Credential Exposure \newline
\textbf{Run2}: Credential \& Secret Exposure \newline
\textbf{Run3}: Credential Exposure \newline
\textbf{Run4}: Secret \& Credential Exposure \newline
\textbf{Run5}: Credential \& Secret Exposure
&
\begin{itemize}
  \item PostgreSQL database passwords should not be disclosed
  \item Google API keys should not be disclosed
  \item Google Cloud service account keys should not be disclosed
  \item GitHub tokens should not be disclosed
  \item GitLab tokens should not be disclosed
  \item Redis credentials should not be disclosed
  \item MySQL database passwords should not be disclosed
  \item Django secret keys should not be disclosed
  \item MongoDB database passwords should not be disclosed
  \item Database passwords should not be disclosed
  \item OpenAI API keys should not be disclosed
  \item Cryptographic private keys should not be disclosed
  \item Credentials should not be hard-coded
\end{itemize}
&
0.82 \\
\hline

\textbf{Run1}: Authentication Issues \newline
\textbf{Run2}: Authentication \& Token Security \newline
\textbf{Run3}: Authentication \& Token Integrity \newline
\textbf{Run4}: Authentication \& Token Security \newline
\textbf{Run5}: Authentication
&
JWT should be signed and verified
&
0.70 \\
\hline

\textbf{Run1}: Lack of Access Control \newline
\textbf{Run2}: Access Control \newline
\textbf{Run3}: Access Control \& Least Privilege \newline
\textbf{Run4}: Authorization \& Privilege Management \newline
\textbf{Run5}: Access Control
&
\begin{itemize}
  \item Service account permissions should be restricted
  \item Wildcards should not be used to define RBAC permissions
  \item AWS IAM policies should limit the scope of permissions given
\end{itemize}
&
0.64 \\
\hline

\textbf{Run1}: CORS Issues \newline
\textbf{Run2}: Cross-Origin Security \newline
\textbf{Run3}: Cross-Origin Security \newline
\textbf{Run4}: Cross-Origin Security \newline
\textbf{Run5}: Cross-Origin Security
&
Origins should be verified during cross-origin communications
&
0.85 \\
\hline

\textbf{Run1}: Improper Resource Management \newline
\textbf{Run2}: Resource Management \newline
\textbf{Run3}: Resource Management \& Hardening \newline
\textbf{Run4}: Resource \& Quota Management \newline
\textbf{Run5}: Resource Management
&
\begin{itemize}
  \item Memory limits should be enforced
  \item Storage limits should be enforced
\end{itemize}
&
0.69 \\
\hline

\textbf{Run1}: Transport Security Issues \newline
\textbf{Run2}: Transport Layer Security \newline
\textbf{Run3}: Transport Layer Security \newline
\textbf{Run4}: Transport Layer Security \newline
\textbf{Run5}: Transport Layer Security
&
\begin{itemize}
  \item Weak SSL/TLS protocols should not be used
  \item Server certificates should be verified during SSL/TLS connections
  \item Server hostnames should be verified during SSL/TLS connections
\end{itemize}
&
0.77 \\
\hline

\textbf{Run1}: Insecure File Creation \newline
\textbf{Run2}: Insecure File Handling \newline
\textbf{Run3}: Insecure Resource Handling \newline
\textbf{Run4}: Secure File Handling \newline
\textbf{Run5}: Insecure File Handling
&
Insecure temporary file creation methods should not be used
&
0.76 \\
\hline

\textbf{Run1}: Input Validation Issues \newline
\textbf{Run2}: Input Validation \newline
\textbf{Run3}: Input Validation \& Deserialization \newline
\textbf{Run4}: Input Validation \& Deserialization \newline
\textbf{Run5}: Input Validation \& Deserialization
&
XML parsers should not be vulnerable to XXE attacks
&
0.79 \\
\hline

\multicolumn{2}{|r|}{\textbf{Average Naming Consistency}} & \textbf{0.75} \\
\hline

\end{longtable}

%
%
%
%
%
%
%
%

\end{document}